\documentclass[useAMS,usenatbib]{mn2e}

\usepackage[pdftex,pdfpagemode={UseOutlines},bookmarks,bookmarksopen,colorlinks,linkcolor={blue},citecolor={blue},urlcolor={red}]{hyperref}
\usepackage{graphicx}	
\usepackage{amsmath}	
\usepackage{amssymb}	
\usepackage{multicol}        
\usepackage{bm}		
\usepackage{pdflscape}	
\usepackage{caption}
\usepackage{subcaption}
\usepackage{natbib}
\usepackage{longtable,lscape}
\usepackage{rotating}
\usepackage{multirow,multicol}

\newcommand{\aap}{A\&A}

\newcommand{\apjs}{ApJS}
\newcommand{\apjl}{ApJL}

\usepackage[T1]{fontenc}
\usepackage{ae,aecompl}

\usepackage{txfonts}

\title[Brown dwarf planets]{First limits on the occurrence rate of short-period planets orbiting brown dwarfs.}

\author[M. He]{Matthias Y. He$^{1,2}$\thanks{Contact e-mail: \href{mailto:matthias.he@mail.utoronto.ca}{matthias.he@mail.utoronto.ca}},
Amaury H.M.J. Triaud$^{3,2,1},$
Micha\"el Gillon$^{4}$\\
$^{1}$Department of Astronomy \& Astrophysics, University of Toronto, Toronto, Ontario, M5S 3H4, Canada\\
$^{2}$Centre for Planetary Sciences, University of Toronto at Scarborough, 1265 Military Trail, Toronto, Ontario, M1C 1A4, Canada\\
$^{3}$Institute of Astronomy, University of Cambridge, Madingley Road, Cambridge, CB3 0H4, UK\\
$^{4}$Institut d'Astrophysique et de G\'eophysique, Universit\'e de Li\`ege, All\'ee du 6 Ao\^ut 17, Sart Tilman, 4000 Li\`ege 1, Belgium\\
}

\date{Accepted ?. Received ?; in original form ?}

\pubyear{2015}

\begin{document}
\label{firstpage}
\pagerange{\pageref{firstpage}--\pageref{lastpage}}
\maketitle

\begin{abstract}
Planet formation theories predict a large but still undetected population of short-period terrestrial planets orbiting brown dwarfs. Should specimens of this population be discovered transiting relatively bright and nearby brown dwarfs, the Jupiter-size and the low luminosity of their hosts would make them exquisite targets for detailed atmospheric characterisation with {\it JWST} and future ground-based facilities. The eventual discovery and detailed study of a significant sample of transiting terrestrial planets orbiting nearby brown dwarfs could prove to be useful not only for comparative exoplanetology but also for astrobiology, by bringing us key information on the physical requirements and timescale for the emergence of life.

In this context, we present a search for transit-signals in archival time-series photometry acquired by the {\it Spitzer Space Telescope} for a sample of 44 nearby brown dwarfs. While these 44 targets were not particularly selected for their brightness, the high precision of their {\it Spitzer} light curves allows us to reach sensitivities below Earth-sized planets for 75\% of the sample and down to Europa-sized planets on the brighter targets. We could not identify any unambiguous planetary signal. Instead, we could compute the first limits on the presence of planets on close-in orbits. We find that within a 1.28 day orbit, the occurrence rate of planets with a radius between 0.75 and 3.25~R$_\oplus$ is $\eta < 67 \pm 1\%$. For planets with radii between 0.75 and 1.25~R$_\oplus$, we place a 95\% confident upper limit of $\eta < 87 \pm 3\%$. If we assume an occurrence rate of $\eta = 27\%$ for these planets with radii between 0.75 and 1.25~R$_\oplus$, as the discoveries of the Kepler-42b and TRAPPIST-1b systems would suggest, we estimate that 175 brown dwarfs need to be monitored in order to guarantee (95\%) at least one detection.




\end{abstract}




\section{Forewords}

Radial velocity and transit surveys have revealed that close-in packed systems of low-mass planets are very frequent around solar-type stars \citep{Mayor:2011fj,Howard:2010zr,Batalha:2013lr} and red dwarfs \citep{Dressing:2013xy}. While still undetected, such systems could also be frequent around brown dwarfs. Observations confirm that brown dwarfs possess the same protoplanetary disc fraction as T\,Tauri stars do \citep{Scholz:2008kx}. In addition \citet{Ricci:2012fk,Ricci:2013lr} found evidence of dust growth, which has also been theoretically explored \citep{Meru:2013yu}. Planet formation is thus expected to occur within these discs \citep{Payne:2007lr}. Despite these encouraging signs, little evidence about short-period planets orbiting brown dwarfs exist, which is despite some of their very advantageous properties for planet detection, but can be explained by their intrinsic faintness.

\begin{figure*}
\centering
\includegraphics[width=0.65\textwidth]{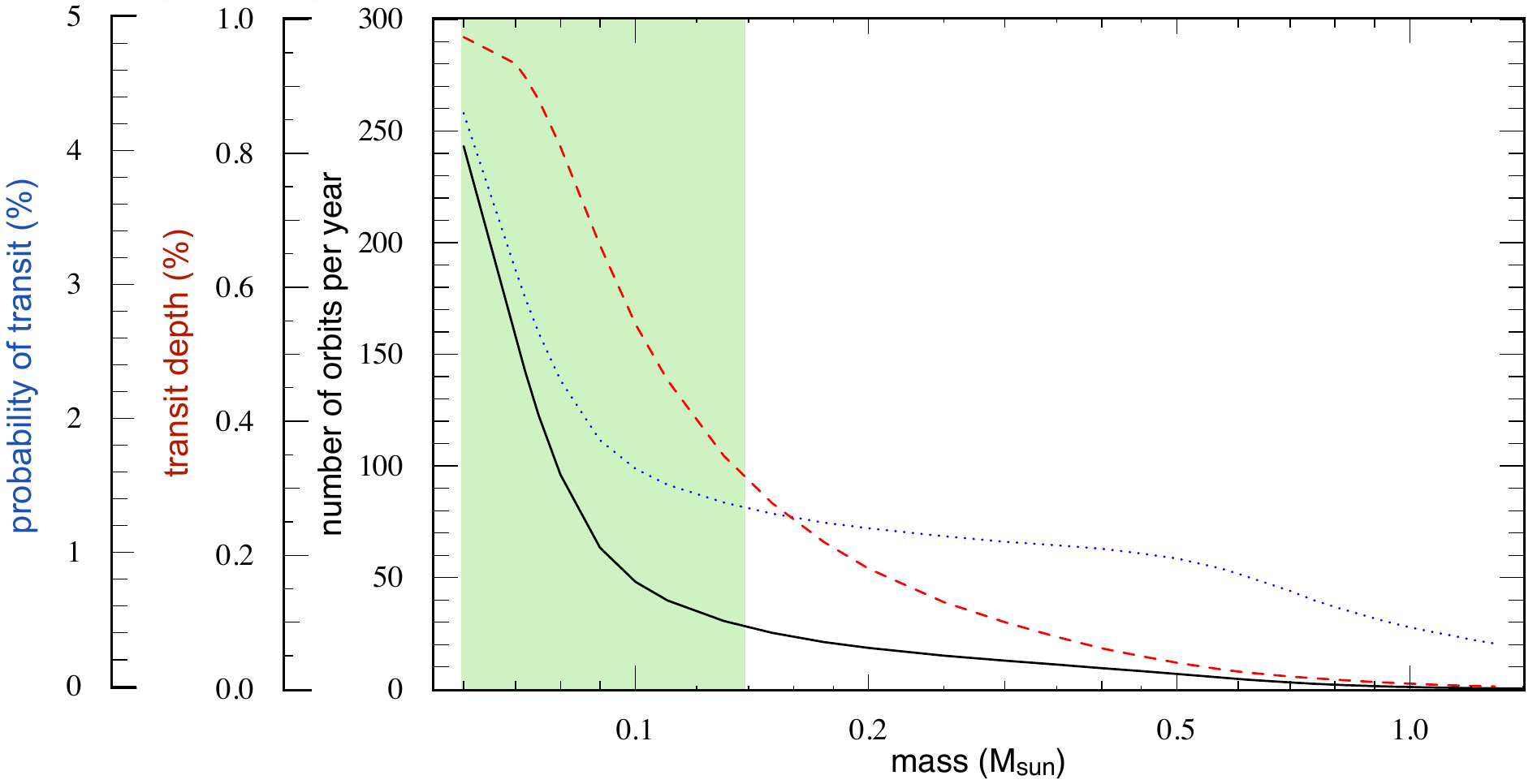}
\caption{For a given equilibrium temperature (here 255K, like Earth), the number of orbits (i.e. transits or occultations) per year (black), the transit depth (dashed red), and the probability of transit (dotted blue) as a function of the primary's mass. The green area is approximatively where a $5\sigma$ on spectral signatures can be reached with the mission lifetime of {\it JWST}. Stellar parameters were obtained from a 1 Gyr isochrone \citep{Baraffe:2003gf}. The slopes steepen with older stellar ages.}
\label{fig:BD}
\end{figure*}

To this day, only a few long-period planetary mass objects have been associated to brown dwarf hosts using direct imaging and microlensing techniques \citep[e.g.][]{Chauvin:2004lr}. The mass ratios and orbital distances however make all these systems resemble substellar binaries more than planetary systems, at the exception of the $\sim 2$ $M_{\rm Jup}$ object OGLE-2012-BLG-0358Lb discovered by \citet{Han:2013qf}. The Carnegie Astrometric Planet Search Program attempts to detect gas giant planets around late M, L, and T dwarfs, as the smaller masses of these stars yield larger astrometric signals for a given planetary companion \citep{Boss:2009lr}. Another astrometric search for long-period planets around brown dwarfs is also currently under way \citep{Sahlmann:2014ys} and has demonstrated sensitivity reaching well into the planetary domain. \citet{Blake:2010yq} used the radial velocity technique and produced a null result. A first attempt to detect short-period terrestrial planets transiting nearby brown dwarfs was performed by \citet{Blake:2007qf}. They used the PAIRITEL infrared telescope to monitor a sample of 20 ultra-cool dwarfs, including some brown dwarfs. This survey did not detect any transiting object, and its precision was too low -- $\sim$1 \% -- with a sample too small to constrain the occurrence of short-period terrestrial planets around brown dwarfs. In 2013, the newly detected nearby binary brown dwarf Luhman-16AB \citep{Luhman:2013qf} was intensively monitored in photometry by the TRAPPIST telescope \citep{Jehin:2011dk,Gillon:2011qf} to search for transiting planets down to the radius of Earth. This project, that failed to detect any transit but revealed the fast-evolving weather of Luhman-16B \citep{Gillon:2013qv}, was done in the context of a transit survey targeting the $\sim$50 brightest Southern ultracool dwarfs ongoing since 2010 on TRAPPIST \citep{Gillon:2013qy}. This same survey identified recently a trio of Earth-sized planets transiting a nearby star with a mass only $\sim$10\% more massive than the Hydrogen-burning limit \citep{gillon:2016gh}. This recent discovery combined with the theoretical prediction that similar systems should be frequent around brown dwarfs is a strong motivation to intensify the search for transiting planets around the nearest brown dwarfs.


In this context, we present here the results of the first space-based search for terrestrial planets transiting brown dwarfs, based on archive data gathered for a sample of 44 brown dwarfs by the {\it Spitzer Space Telescope}. In the next section we outline the importance of a search for planets transiting brown dwarfs. In Section~\ref{sec:sample} we describe the sample we use and in Sect.~\ref{sec:signal} perform early calculations on what type of planets can be detected. We then present a search algorithm in Section~\ref{sec:retrieval}, and test it using synthetically inserted transits. We apply this algorithm to first seek planets within the sample (Sect.~\ref{sec:search}) and then use it to compute upper limits on the occurence rate (Sect.~\ref{sec:occurence}). We then conclude.

\section{A case for finding planets transiting brown dwarfs}

We briefly summarise here a case that we presented in a white paper  \citep{Triaud:2013lh} and which we detailed in a number of observing proposals, attempting to monitor brown dwarfs in search for systems of transiting planets. Some of our arguments are similar to those made in favour of M dwarfs by \citet{Nutzman:2008qy}, particularly late M dwarfs \citep{Kaltenegger:2009qf,Belu:2011qy,Rodler:2014zl}. 

Brown dwarfs have characteristics that make them ideal targets to search for Earth-like rocky worlds, but also optimal for their atmospheric characterisation. Two hundred brown dwarfs have near-infrared magnitudes $K< 13$ (\href{http://www.dwarfarchives.org}{dwarfarchives.org}), which are optimal for {\it JWST}. As an example, a first attempt to perform transmission spectroscopy on the TRAPPIST-1b \& 1c planets has been presented by \citep{de-Wit:2016fk} using the {\it Hubble} space telescope. Any planets found transiting a brown dwarf will offer similarly good conditions for the {\it JWST}, if not more, on account of their smaller radii, which enhance transmission features.


Emission and reflection spectroscopy, as well as phase curves, will complement transmission spectroscopy and provide additional information about the atmospheres of any discovered exoplanet \citep{Seager:2010kx}, as will high-resolution spectroscopy \citep{Collier-Cameron:1999qf,Snellen:2015kq}. A planet the size and temperature of the Earth has the same blackbody emission be it orbiting a G dwarf or a brown dwarf. The latter however is 6 to 10 magnitudes fainter, a very favourable case akin to a {\it natural coronograph}. Emission and reflection spectroscopy probe the full face of the planet, with flux emerging through only one airmass, which makes the technique less sensitive to clouds than transmission spectroscopy. In addition, occultations permit the elaboration of thermal maps \citep{de-Wit:2012uq}. Measuring the reflected and emission spectra can also be obtained when the system is not in a transiting configuration \citep{Snellen:2015kq}. In some configurations a 255~K, Earth-sized planet can have a signal of order that achieved these days for hot Jupiters orbiting by G dwarfs \citep[$\sim 1e^{-5}$, e.g.][]{Leigh:2003yu,Brogi:2014fv}. 

Other advantages are represented graphically in Figure~\ref{fig:BD}, where we show two metrics of detection (probability of transit and transit depth), and one metric of characterisation (number of orbits per year), for an Earth-sized planet, with an equilibrium temperature of 255 K (like Earth), as a function of stellar mass. By focusing on brown dwarfs, we gain one order of magnitude on the probability of having a habitable zone planet in a transiting configuration \citep{Bolmont:2011uq}. We also improve by two orders of magnitude the depth of the transit, easing detection, and reaching a comfortable 1\% transit depth, which is routinely detected by many ground-based instruments \citep[][]{Pollacco:2006fj,Bakos:2007fj,Gillon:2013qv,gillon:2016gh}. Finally, for a true Earth analog orbiting a solar analog, there is only one transit per year. If 15 transits need to be co-added to reach a significant detection of atmospheric features, 15 years of observations are required. For brown dwarfs, we can collect in excess of one transit, and one occultation, every week. 

\citet{Belu:2011qy} point out that a formal detection of atmospheric features can only be made during the limited mission lifetime of {\it JWST}, if that planet orbits a star with a mass inferior to 0.2~M$_\odot$. This is assuming that all transit events will be recorded. Due to scheduling concerns, it is more likely closer to 0.15~M$_\odot$. Brown dwarfs form a significant fraction of the remaining systems. Any planet (rocky or not, habitable or not), found to be transiting a nearby brown dwarf could make an exquisite target for a detailed atmospheric characterisation.

\section{Our sample}\label{sec:sample}
\begin{figure*}
\centering
\includegraphics[width=0.85\textwidth]{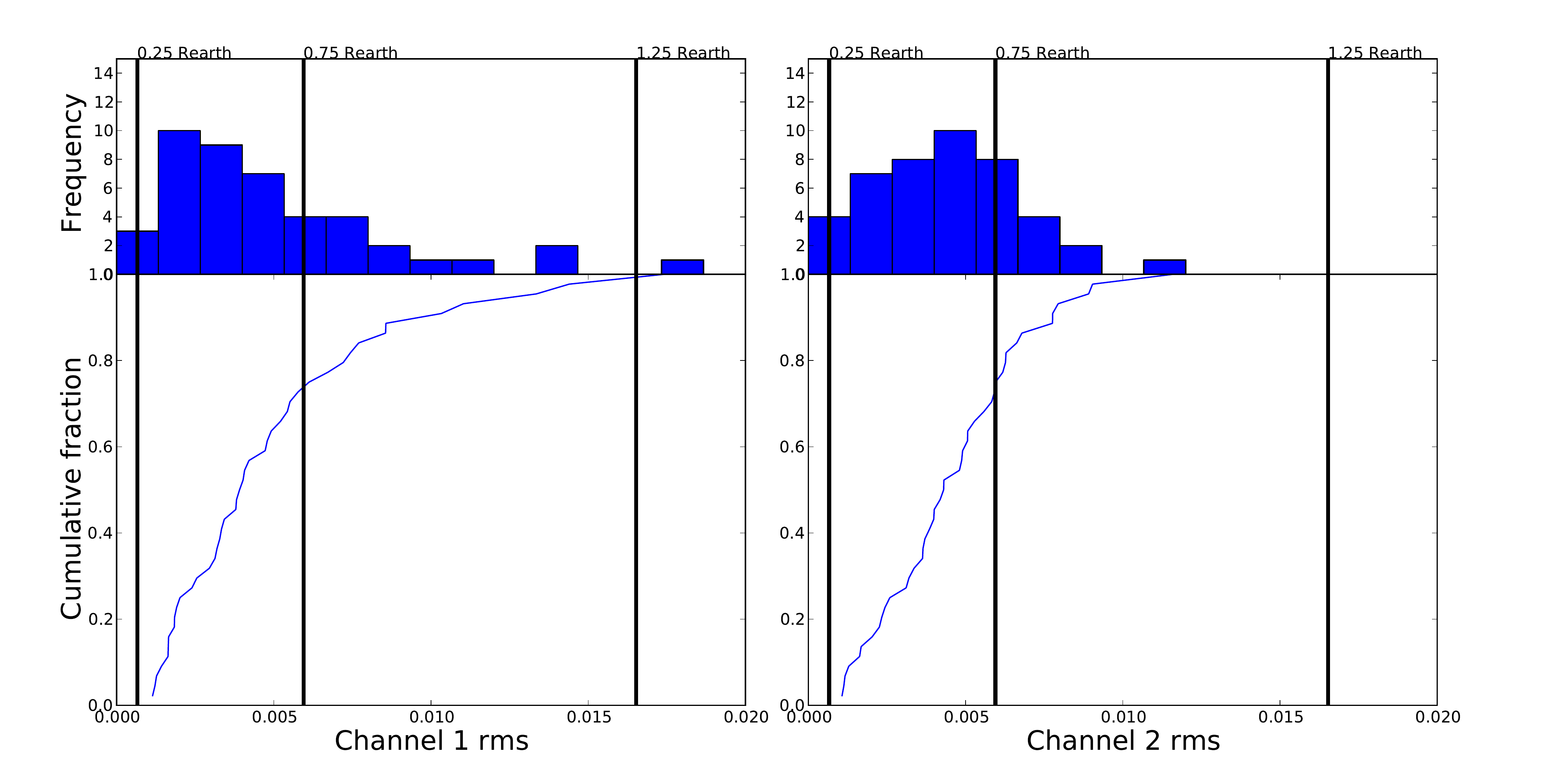}
\caption{Histogram and cumulative distributions of the rms of the {\it Spitzer} photometry on our sample of 44 brown dwarfs, in both channels. Vertical lines translate the rms into detectable planetary radii, assuming transits with a signal to noise ratio of at least 3.5.}
\label{fig:rmshistograms}
\end{figure*}

\begin{table}
\centering
\caption{Examples of typically produced transit depths and transit widths $W$, with their corresponding signal to noise ratio (in brackets). We assumed a 1$R_\oplus$ planet, an impact parameter of $b = 0.5$, and a point to point rms of $\sigma = 0.49\%$ (taken as the weighted average of both channels for all the objects).}
\label{tab:example}
\tiny
\resizebox{0.49\textwidth}{!}{
\begin{tabular}{cccccc}
\hline
\hline
M$_\star$ 		& R$_\star$ 		&$D$	& W ($P = 0.23$d)	&W ($P =  0.55$d)	& W ($P =  1.28$d)\\
{[ M$_{\rm Jup}$ ]}	&  {[ R$_{\rm Jup}$ ]}	&{[ $\%$ ]}	&	[min]	 (SNR)	&	[min]	 (SNR)	& 	[min] (SNR)	\\
\hline
30 				& 0.8 			& 1.30 	& 17.5 (7.5)		& 23.3 (8.7)		& 30.9 (10.0)		\\
30 				& 0.9 			& 1.03 	& 19.4 (6.2) 		& 25.9 (7.2)		& 34.3 (8.3)		\\ \vspace{0.5em}
30 				& 1.0 			& 0.83 	& 21.3 (5.3) 		& 28.4 (6.1) 		& 37.7 (7.1)		\\
60 				& 0.8 			& 1.30 	& 13.9 (6.7) 		& 18.5 (7.7) 		& 24.6 (8.9)		\\
60 				& 0.9 			& 1.03 	& 15.4 (5.6) 		& 20.5 (6.4) 		& 27.2 (7.4)		\\ \vspace{0.5em}
60 				& 1.0 			& 0.83 	& 16.9 (4.7) 		& 22.6 (5.5) 		& 29.9 (6.3)		\\
80 				& 0.8 			& 1.30 	& 12.6 (6.4) 		& 16.8 (7.4) 		& 22.3 (8.5)		\\
80 				& 0.9 			& 1.03 	& 14.0 (5.3) 		& 18.7 (6.1) 		& 24.7 (7.1)		\\
80 				& 1.0 			& 0.83 	& 15.3 (4.5) 		& 20.5 (5.2) 		& 27.2 (6.0)		\\
\hline
\end{tabular}}
\end{table}

The only sizeable sample of brown dwarfs that has been monitored photometrically in a sufficiently precise and consistent manner, has been presented by \citet{Metchev:2015rf}. They acquired their data using the {\it Spitzer} Space Telescope (in its {\it warm} phase), as part of an Exploration Science program called {\it Weather on Other Worlds} \citep[see also][]{Heinze:2013lr}. Their aim was to study the mid-infrared variability of brown dwarfs across the L to T spectral class transition. 44 brown dwarfs ranging from L3 to T8 were monitored using IRAC \citep{Fazio:2004fk}, in both channel 1 and 2 (3.6~$\mu$m and 4.5~$\mu$m respectively, or [3.6] \& [4.5]). The main results are presented in \citet{Metchev:2015rf}. Out of 44 brown dwarfs, 23 exhibit constant photometric timeseries  while the other 21 display significant variability. Variability is produced by any inhomogeneity on the surface of brown dwarfs that moves in and out of view due to rotation.
 \citet{Metchev:2015rf} provide arguments implying that all brown dwarfs may in fact be variable. Those that have their rotation axis aligned with the line of sight, present a similar hemisphere at any given time, and produce constant timeseries. The physical origin of this variability remains debated but is now frequently interpreted as being produced by clouds \cite[e.g.][]{Ackerman:2001fk,Burgasser:2002kx,Artigau:2009lr,Radigan:2012rt,Gillon:2013qv,Radigan:2014lr,Crossfield:2014db}.
 
For our analysis, we used the data already detrended from rotational variability that is presented in \citet{Metchev:2015rf}. In order to study variability and reach their conclusions, \citet{Metchev:2015rf} fit Fourier components to their data. Most often, one term is clearly significant. When several alternate models are proposed, we adopt the model according to the following criteria: we resorted to the lowest number of Fourier terms unless each extra term improved the point to point rms by more than 10\%, or significant variability could still be seen after correction using the lowest number of terms. We now have a set of 44 photometric timeseries that has been corrected for both instrumental and astrophysical effects. We search for transits into each lightcurve and use the set to place the first constraints on the presence of  planets orbiting brown dwarfs.

The median observation time per brown dwarf is 20.9 hours, over both channels. Typically, observations start with the 3.6 $\mu$m channel, observing for 13.8h (exceptions are 2M0036, 2M1632, and DENIS1058, which only have about 7-8h) and then proceed to 4.5 $\mu$m for 6.8h (except 2M2224, 2M0825, and 2M1507, which have 8-9h of data). There is a small gap while the spacecraft moves from channel 1 to channel 2. The median gap between is 15 minutes, although 2M0949 and 2M1511 have gaps that are over an hour long (1.9h and 1.4h respectively). All targets, in both channels, exhibit a similar cadence, with new frames acquired every 2.18 min $\pm$ 0.07 min (the error is taken as standard deviation of the gaps in time between consecutive data points, ignoring the larger gap between the two channels). Statistically, the two photometric channels produce a similar precision: on average, the flux at [3.6] has a point to point rms of 0.0050 $\pm$ 0.0037 while [4.5] has an rms of 0.0046 $\pm$ 0.0024  (the errors are taken as the standard deviations of the rms over all the objects, per channel). We plot their distribution in Fig.~\ref{fig:rmshistograms}. 

We note a difference between the rms of the objects with constant light curves and that of the variable objects. Constant timeseries have a higher dispersion than variable targets. On average constant timeseries have an rms =  0.0060 $\pm$ 0.0032 while this value is 0.0035 $\pm$ 0.0023 for those light curves with variability removed. We do not elaborate much on this topic, but remark that this may suggest that the Fourier treatment overfits the data. Alternatively, it is possible that some {\it constant} sources are in fact {\it variable}, but that their variability is too weak to be significant, or that it is not coherent in time \citep[e.g.][]{Artigau:2009lr,Gillon:2013qv}. 

\section{Expected signal}\label{sec:signal}

\begin{figure*}
\begin{center}
\begin{subfigure}[b]{0.45\textwidth}
	\includegraphics[trim={0 0.9cm 0 0},clip=true,scale=0.4,width=\textwidth]{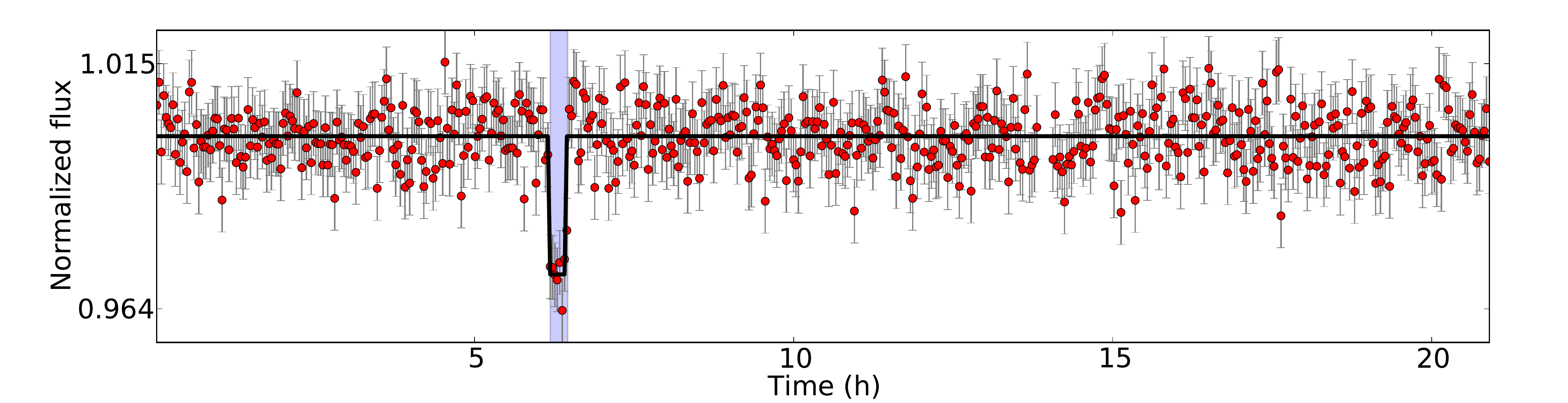}
	\includegraphics[trim={0 0.9cm 0 0},clip=true,scale=0.4,width=\textwidth]{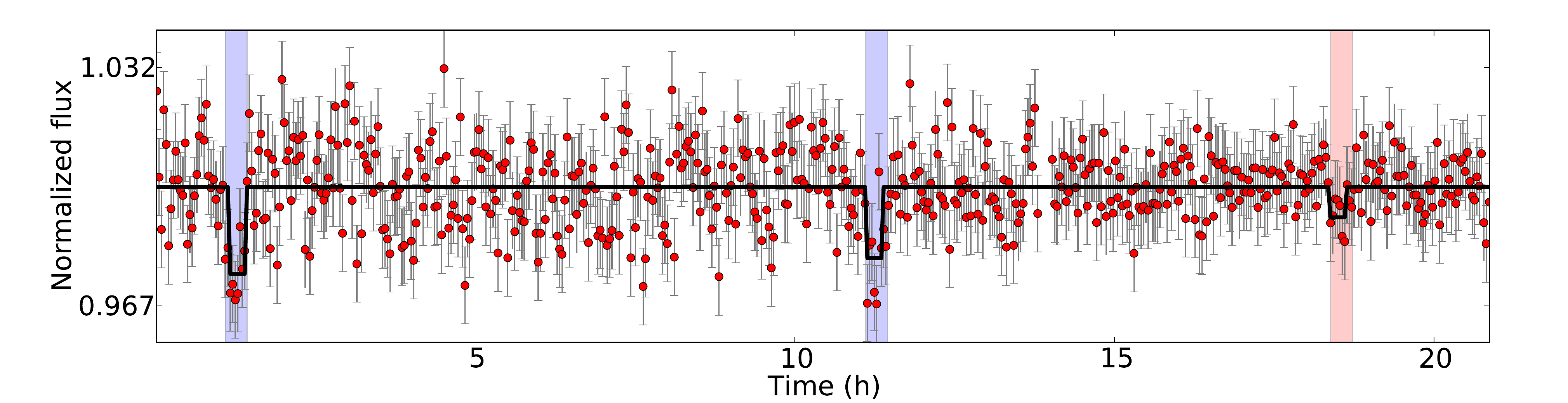}
	\includegraphics[trim={0 0.9cm 0 0},clip=true,scale=0.4,width=\textwidth]{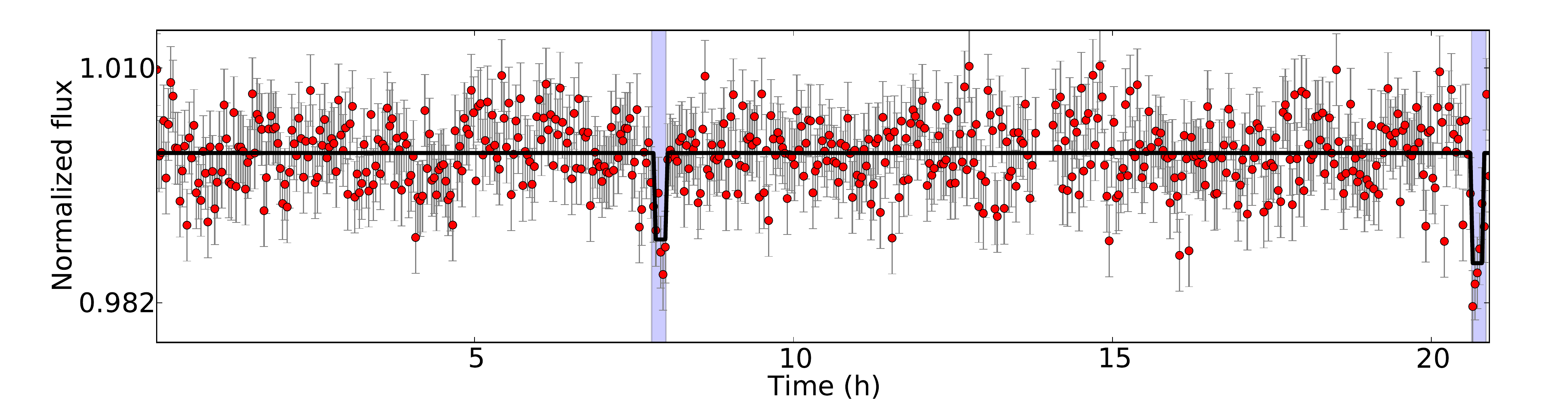}
	\includegraphics[scale=0.4,width=\textwidth]{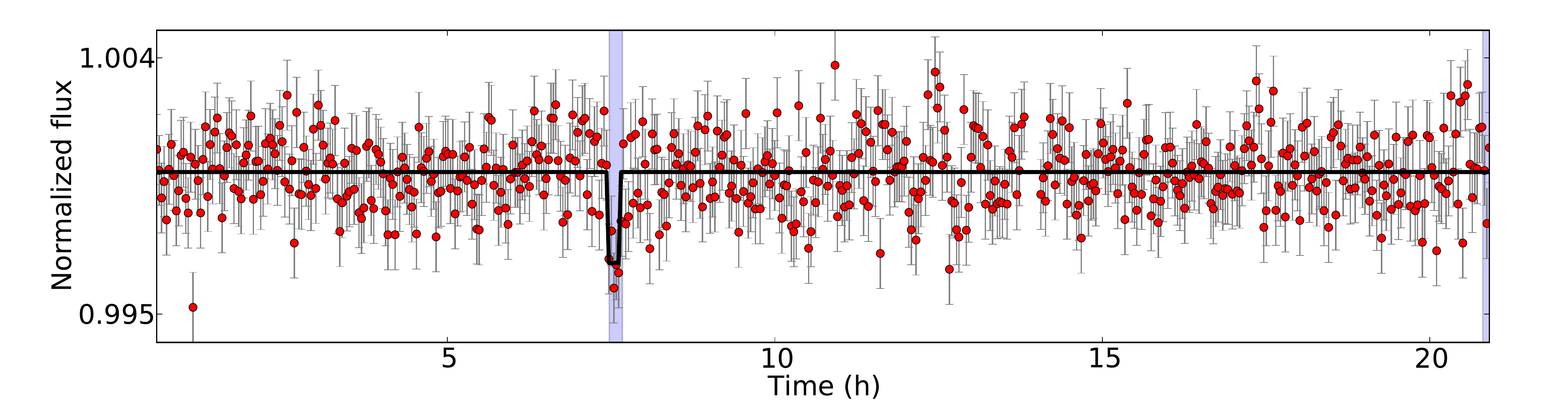}
\end{subfigure}
\begin{subfigure}[b]{0.45\textwidth}
	\includegraphics[trim={0 0.9cm 0 0},clip=true,scale=0.41,width=1.06\textwidth]{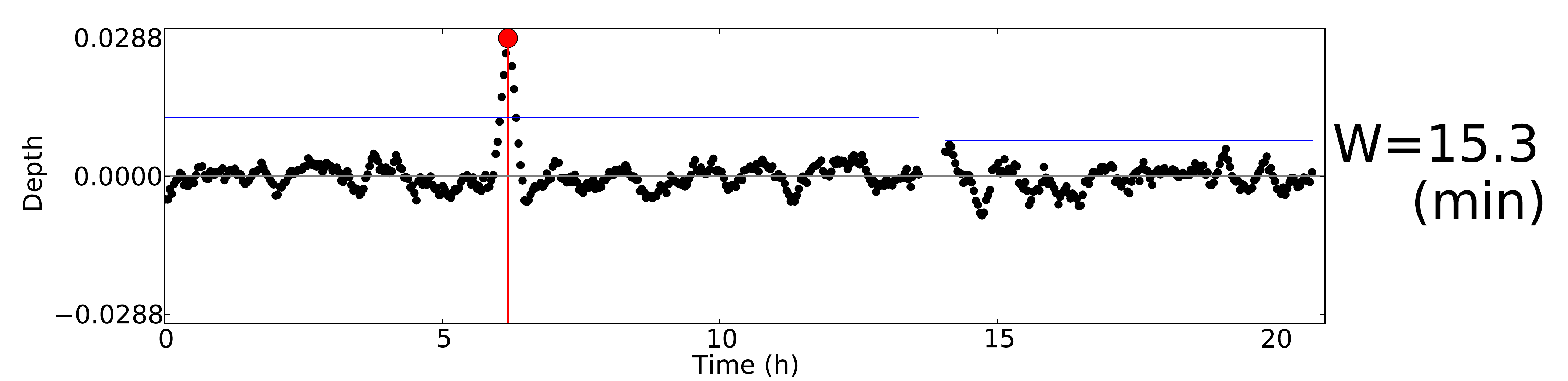}
	\includegraphics[trim={0 0.9cm 0 0},clip=true,scale=0.41,width=1.06\textwidth]{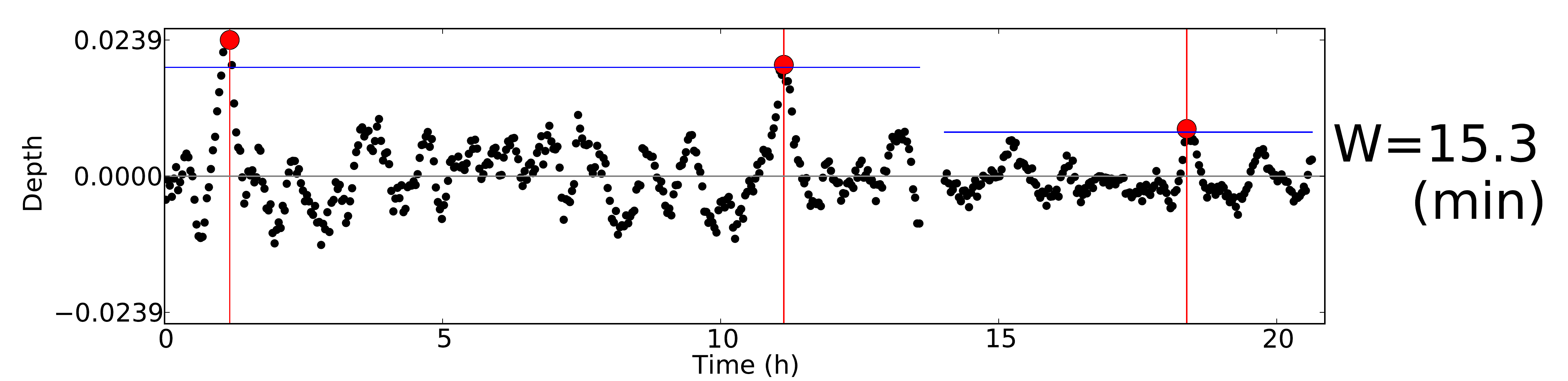}
	\includegraphics[trim={0 0.9cm 0 0},clip=true,scale=0.41,width=1.06\textwidth]{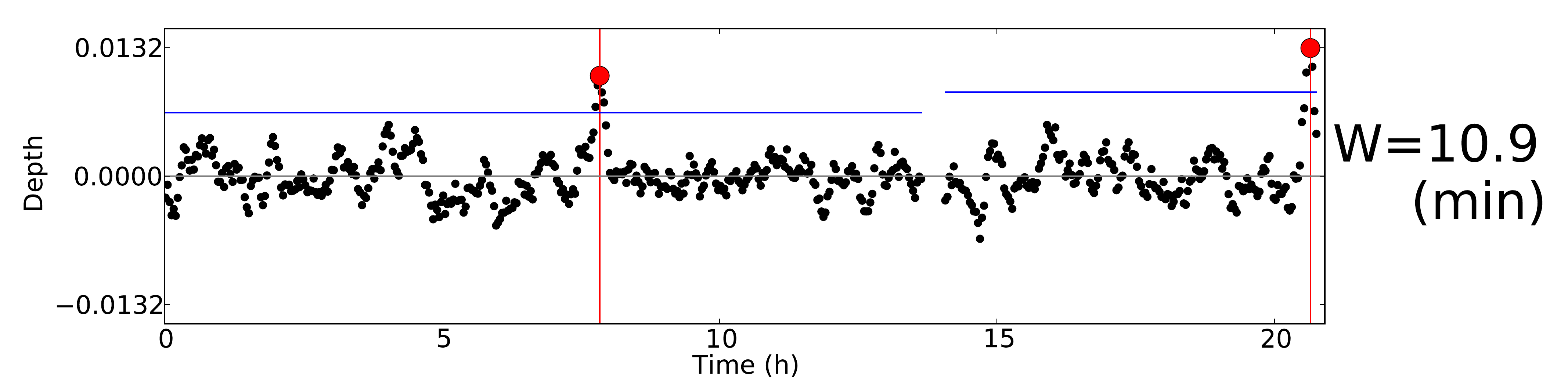}
	\includegraphics[scale=0.41,width=1.06\textwidth]{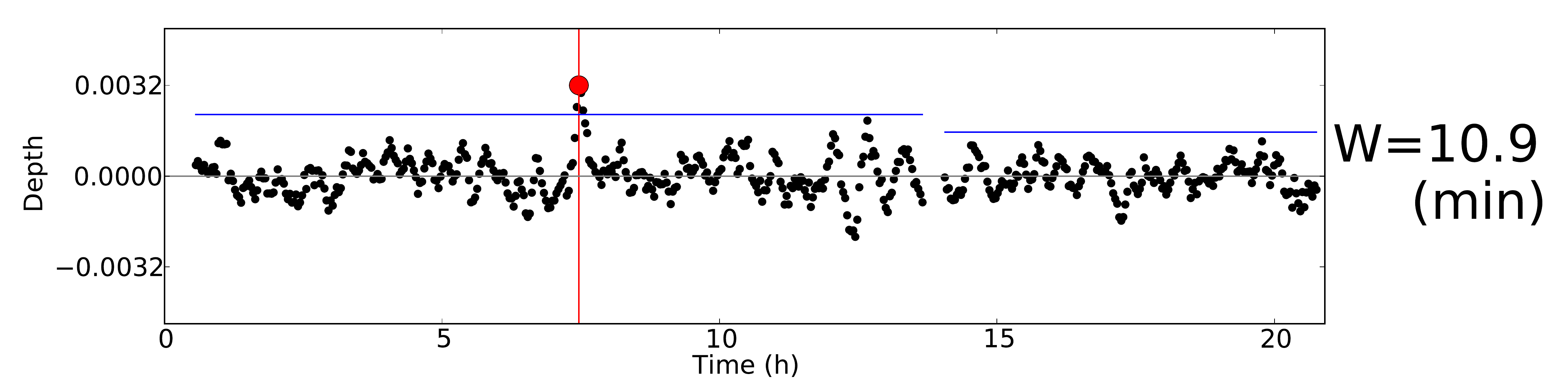}
\end{subfigure}
\caption{A few examples of our ability to retrieve planetary transits that we added to the {\it Spitzer} lightcurves (locations marked with a small vertical line). {\it Left}, we have four brown dwarfs' lightcurves, as observed with {\it Spitzer} showing artificially implanted planetary transits. {\it Right}, we show the depth detected by our top-hat model, as a function of time for each of this timeseries. We also display the mean and three times the rms above the mean, with solid blue lines. Large red dots with accompanying vertical red lines indicate where transit-like signals are identified. From top to bottom: (a) Lightcurve of SDSS 2052 containing a correct retrieval of an inserted transit produced by a planet of 1.75~R$_\oplus$. (b) This is 2MASS 0516, with a 1.41~R$_\oplus$ planet transiting twice; both instances are correctly recovered, but a third, false positive signal too (indicated by the red stripe). (c) We have two simulated transits of a 1.0~R$_\oplus$ planet on the time series of SDSS 1520, both of which were easily retrieved. (d) Shows 2MASS 2148, where two transits were artificially inserted, to simulate a 0.59~R$_\oplus$ planet on a short orbit; we identify only one of the transits, failing to recover the second, a false negative. We think the second event is missed by the search algorithm due to it being barely at the very end of the time series.}
\label{fig:lightcurvesmodels}
\end{center}
\end{figure*}

The important variables used to analyse planetary transits are described by  \citet{Mandel:2002kx,Seager:2003qy} and \citet{Winn:2010lz}. The mass of the brown dwarf, the orbital period of the planet, the relative size of the brown dwarf and the planet, as well as the orbital inclination on the sky, essentially produce two observables, the depth $D$, and the width $W$. The depth is the ratio of areas squared, $D = R^2_{\rm p} / R^2_\star$, for $R_{\rm p}$, the planet's radius, and $R_\star$, a brown dwarf's.

We assume a fiducial brown dwarf for the remainder of the paper. Its mass is $M_\star = 60 M_{\rm Jup}$ and its size is $R_\star = 0.9 R_{\rm Jup}$. Brown dwarfs $> 1$ Gyr all have approximately this size \citep[e.g.][]{Baraffe:2003gf,Triaud:2013lr,Moutou:2013fk,Diaz:2013fr,Dieterich:2014lr}, which eases greatly how we sample parameter space. Further justifications for these numbers will come through the paper. 
However what we assume for the brown dwarf does not affect our conclusions much. Furthermore, our retrieval only uses $D$ and $W$ as observables, which are independent of what we assume for the central object. In Table~\ref{tab:example}, we produce a series of typical depth and width values, for a few assumed brown dwarf parameters. Since the cadence of observation is the same for each series, we will also frequently refer to the width as a {\it number of points in transit}, or $n$.

We now assess how good the {\it Spitzer} lightcurves are in terms of what type of planets they are be able to yield. For this purpose, we define a signal to noise ratio (SNR) such that:
\begin{equation}\label{eq:snr}
SNR = \sqrt{n} {D \over \sigma}
\end{equation}
where $n$ is the number of data points that fall within transit, $D$ is the transit depth, our signal, and $\sigma$ is the point to point rms of the timeseries. A typical transit (Table~\ref{tab:example}) has a width $W$ that lasts 22.2 minutes; we chose here $n=10$ (meaning a width of 22 minutes). For a set SNR, we can calculate approximately which planetary radii $R_{\rm p}$ would be detected for a certain rms. For reasons elaborated in Section~\ref{sec:retrieval} and shown in Fig~\ref{fig:SNRhistograms}, we set SNR = 3.5. We draw the histogram  and cumulative distributions of the rms of our sample in Fig.~\ref{fig:rmshistograms}, and indicate which rms can allow the identification of planets with 0.25, 0.75 and 1.25~R$_\oplus$. For roughly 75\% of the sample, we are sensitive to planets with $R_{\rm p}$ down to 0.75 R$_\oplus$, which is remarkable.

\section{Retrieval of artificially inserted transits}\label{sec:retrieval}

The {\it Spitzer} timeseries that we have at our disposal are relatively short (20.9 hours). This means that we need a detection algorithm able to retrieve single transit-like signals.  Because our sample is also small, we can allow ourselves fairly loose detection criteria. The systems that are picked up can be analysed in detail and could lead, in a second stage, to follow-up observations whose aim would be the confirmation of a transit signal, the measure of the planet's period, as well as possibly, evidence for the presence of other planetary companions.

\begin{figure}
\centering
\begin{subfigure}{0.45\textwidth}\caption{2M2254}
	\includegraphics[width=\textwidth]{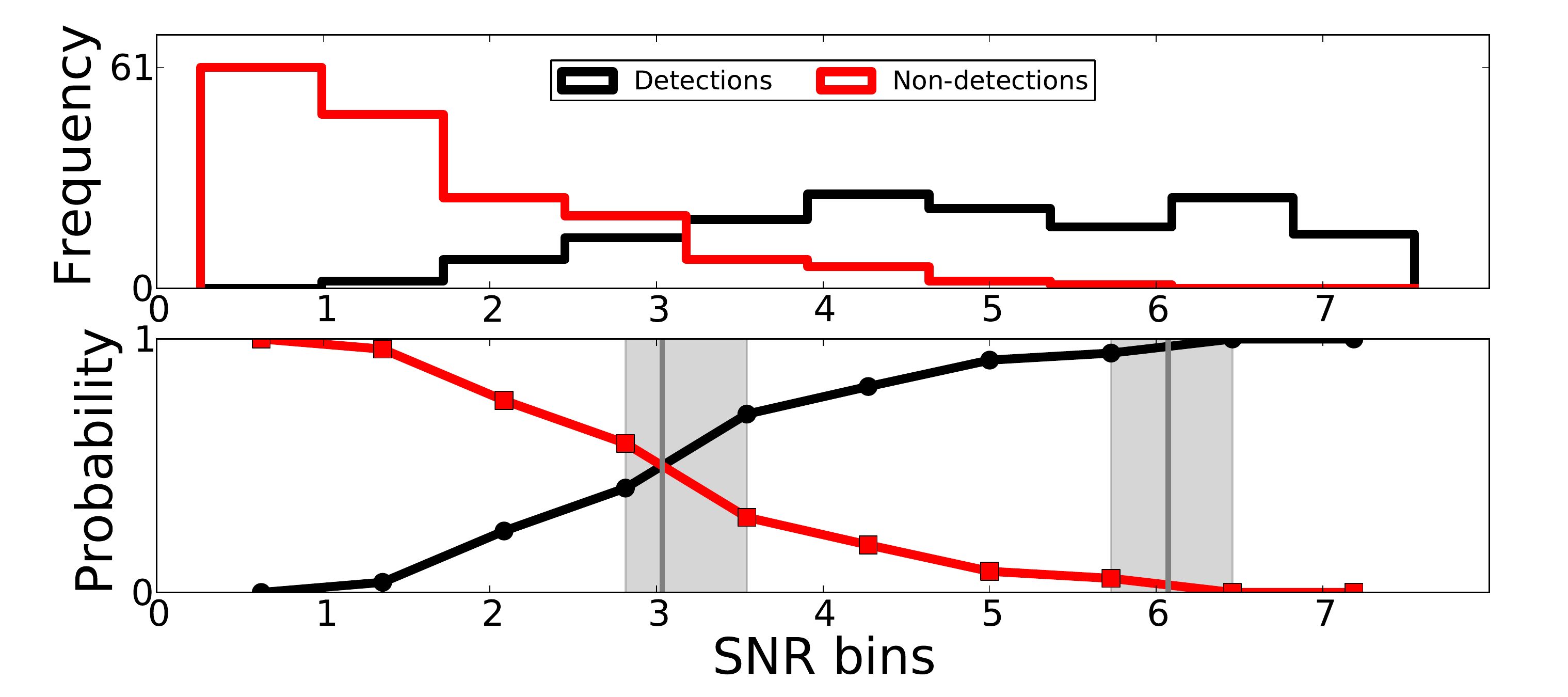}
\end{subfigure}
\begin{subfigure}{0.45\textwidth}\caption{2M0103}
	\includegraphics[width=\textwidth]{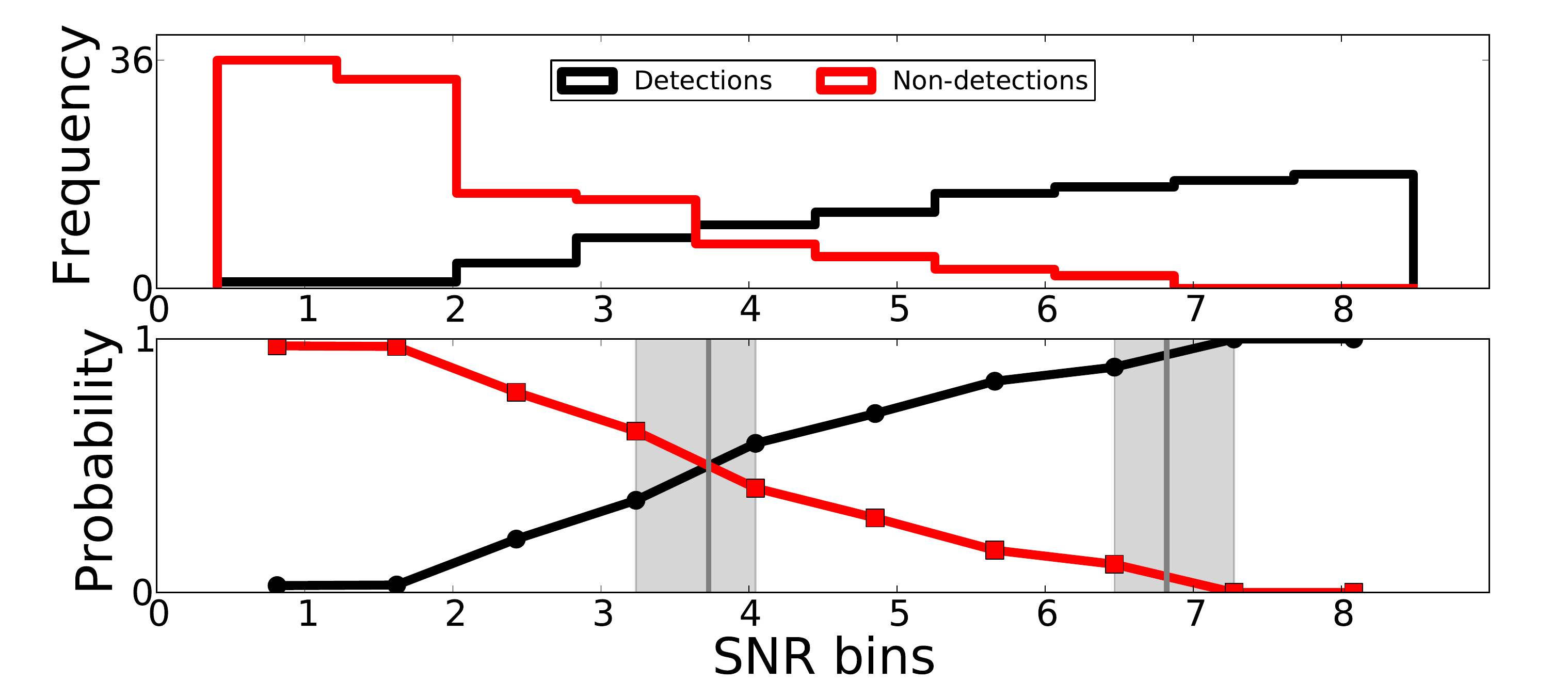}
\end{subfigure}
\caption{Examples of our recovery rate for two brown dwarfs. Detections and non-detections are plotted in black and red, respectively. {\it Top}, we plot how systems distribute in bins of signal to noise ratios. {\it Bottom}, we draw the respective fraction of recovered to non recovered for each of the bins. For clarity, only ten bins were used for each histogram, and include all the inserted transits that have an SNR less than the highest SNR of the undetected transits plus one. The SNR corresponding to the 50\% and 95\% recoverability rates (represented by vertical black lines) were calculated by linear interpolation between the bins (grey regions).}
\label{fig:SNRhistograms}
\end{figure}

\begin{table}
\centering
\caption{Signal to noise ratios for the time series of each object corresponding to a 50\% and 95\% chance of transit retrieval, as well as the maximum SNR out of all assumed transit widths for each object. Their respective planet radii were calculated from adopting an orbital period equal to the average coverage of the time series (20.9h) and an impact parameter $b = 0.5$. The maximum SNR marked by * correspond to signals that are both detected by our 3 rms criterion and over the 50\% SNR threshold.}
\label{tab:SNR50and95}
\begin{tabular}{l|c|c|c|c|c}
\hline
\hline
Object & SNR & R$_p$ & SNR & R$_p$ & Max. \\
 & for 50\% & (R$_\oplus$) & for 95\% & (R$_\oplus$) & SNR \\
\hline
2MASS 0036 & 5.4 & 0.33 & 8.9 & 0.43 & 4.7 \\
2MASS 0050 & 3.6 & 0.85 & 5.8 & 1.08 & 3.4 \\
2MASS 0103 & 3.7 & 0.48 & 6.8 & 0.65 & 3.7 \\
SDSS 0107  & 3.5 & 0.42 & 6.4 & 0.57 & 3.7 \\
SDSS 0151  & 3.5 & 0.69 & 5.5 & 0.87 & 3.3 \\
2MASS 0328 & 3.0 & 0.57 & 6.7 & 0.85 & 4.0* \\
2MASS 0421 & 3.5 & 0.42 & 7.4 & 0.60 & 3.1 \\
2MASS 0516 & 4.1 & 0.86 & 5.1 & 0.95 & 3.3 \\
2MASS 0820 & 3.3 & 0.52 & 4.7 & 0.63 & 3.1 \\
2MASS 0825 & 3.6 & 0.37 & 6.1 & 0.48 & 3.0 \\
SDSS 0858  & 3.1 & 0.55 & 6.0 & 0.76 & 2.5 \\
2MASS 0949 & 3.0 & 0.68 & 5.6 & 0.92 & 3.1 \\
SDSS 1043  & 4.8 & 0.65 & 8.8 & 0.89 & 5.0 \\
DENIS 1058 & 4.0 & 0.40 & 6.8 & 0.51 & 2.9 \\
2MASS 1059 & 2.9 & 0.85 & 4.6 & 1.06 & 3.0 \\
2MASS 1122 & 3.1 & 0.48 & 5.8 & 0.66 & 3.2 \\
2MASS 1126 & 3.5 & 0.38 & 6.3 & 0.51 & 3.1 \\
SDSS 1150  & 3.2 & 0.55 & 6.1 & 0.76 & 3.1 \\
2MASS 1209 & 2.8 & 0.59 & 5.2 & 0.79 & 3.3* \\
SDSS 1254  & 3.2 & 0.44 & 4.6 & 0.53 & 3.0 \\
ROSS 458C  & 4.1 & 1.04 & 6.8 & 1.34 & 3.4 \\
2MASS 1324 & 3.3 & 0.44 & 4.9 & 0.54 & 3.0 \\
ULAS 1416  & 4.3 & 0.81 & 6.4 & 0.99 & 3.4 \\
SDSS 1416  & 3.7 & 0.34 & 6.7 & 0.45 & 3.7 \\
2MASS 1507 & 3.9 & 0.31 & 7.4 & 0.43 & 3.8 \\
SDSS 1511 & 3.2 & 0.51 & 6.6 & 0.73 & 2.7 \\
SDSS 1516 & 2.8 & 0.51 & 4.7 & 0.67 & 4.0* \\
SDSS 1520  & 3.8 & 0.53 & 6.1 & 0.67 & 4.1* \\
SDSS 1545  & 3.9 & 0.75 & 6.2 & 0.96 & 3.9 \\
2MASS 1615 & 3.0 & 0.44 & 7.9 & 0.71 & 2.9 \\
2MASS 1632 & 4.7 & 0.57 & 8.5 & 0.76 & 3.6 \\
2MASS 1721 & 2.6 & 0.31 & 4.3 & 0.40 & 3.5 \\
2MASS 1726 & 3.3 & 0.46 & 6.8 & 0.66 & 3.1 \\
2MASS 2224 & 3.4 & 0.32 & 5.5 & 0.41 & 3.1 \\
2MASS 1753 & 3.3 & 0.33 & 6.5 & 0.46 & 2.9 \\
2MASS 1821 & 5.4 & 0.35 & 8.0 & 0.43 & 4.8 \\
SDSS 2043  & 3.5 & 0.72 & 4.8 & 0.85 & 4.4 \\
SDSS 2052  & 3.3 & 0.59 & 6.4 & 0.82 & 2.9 \\
HN\,Peg\,B     & 2.9 & 0.56 & 6.0 & 0.80 & 2.8 \\
2MASS 2148 & 3.4 & 0.27 & 7.0 & 0.39 & 3.3 \\
2MASS 2208 & 3.0 & 0.46 & 4.5 & 0.57 & 3.7* \\
2MASS 2228 & 4.2 & 0.77 & 6.3 & 0.93 & 4.8* \\
SDSS 2249  & 3.0 & 0.48 & 5.6 & 0.66 & 2.5 \\
2MASS 2254 & 3.0 & 0.57 & 6.1 & 0.80 & 3.3 \\
\hline
\end{tabular}
\end{table}

\begin{figure*}
\centering
\begin{subfigure}{.44\textwidth}\caption{2M0328}
\centering
\includegraphics[trim={0.2cm 1.3cm 0 0.95cm},clip=true,width=\linewidth]{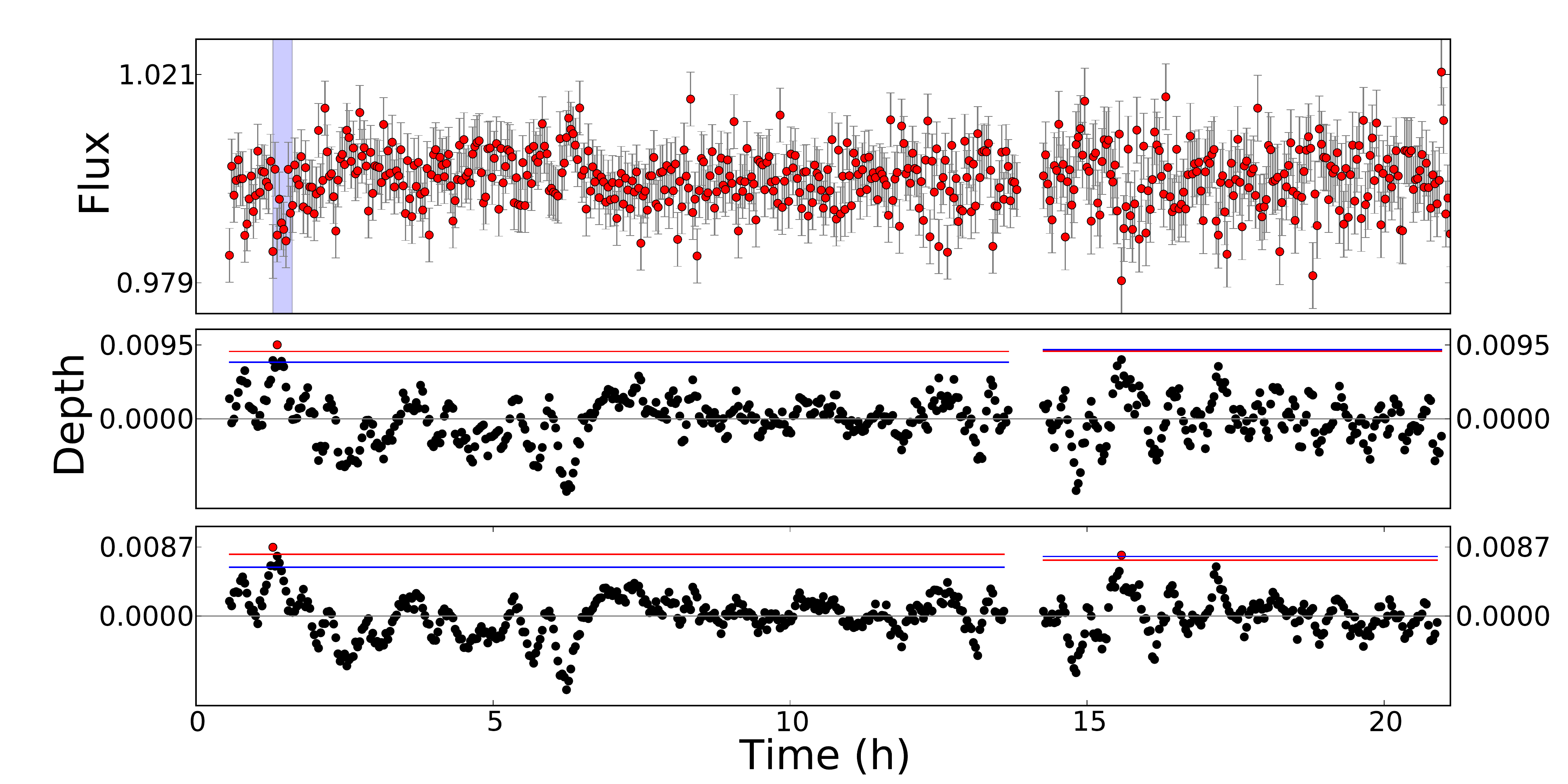}
\end{subfigure}
\begin{subfigure}{.44\textwidth}\caption{2M1209}
\centering
\includegraphics[trim={0.2cm 1.3cm 0 0.95cm},clip=true,width=\linewidth]{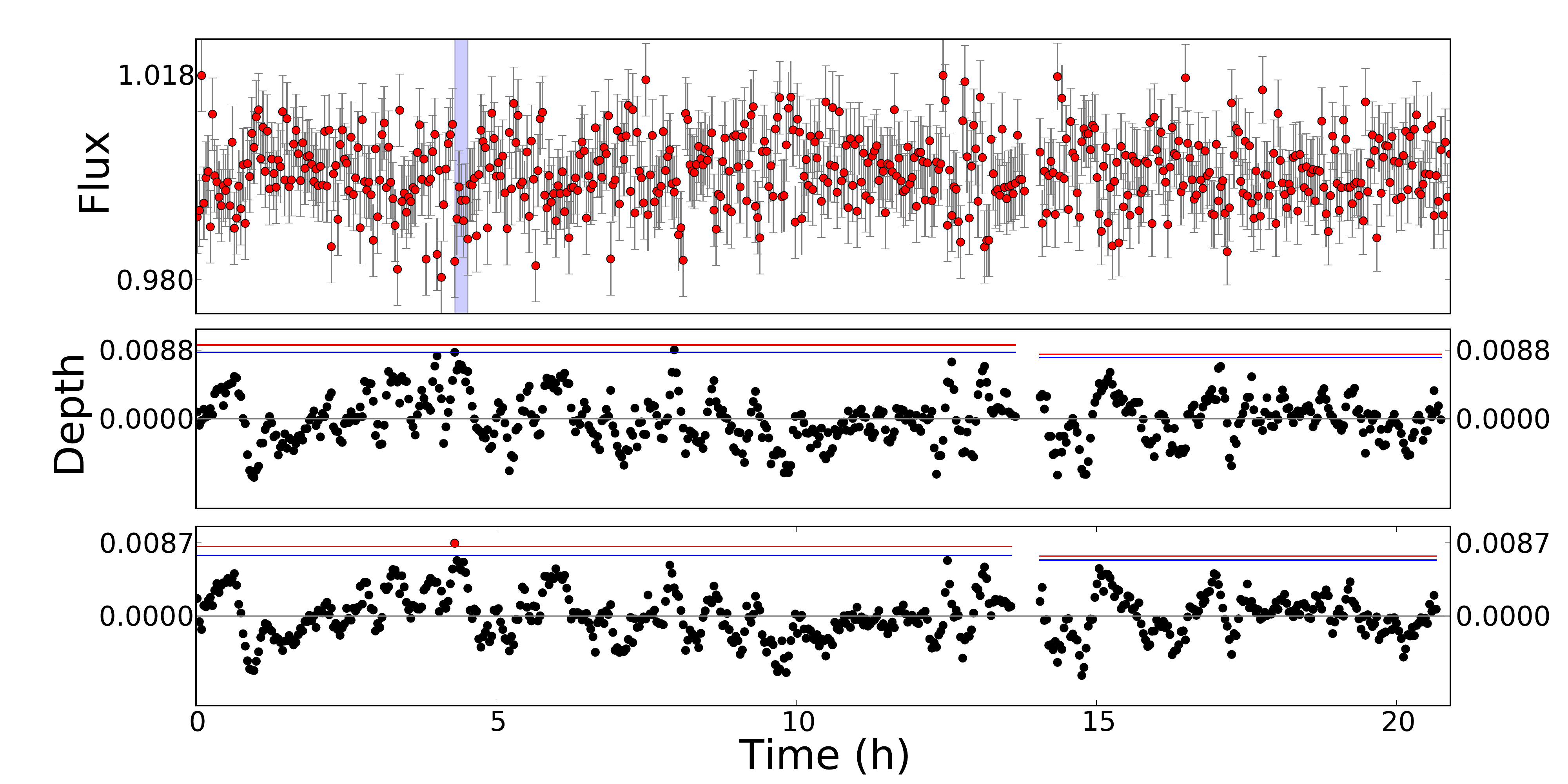}
\end{subfigure}
\begin{subfigure}{.44\textwidth}\caption{SDSS1520}
\centering
\includegraphics[trim={0.2cm 1.3cm 0 0.95cm},clip=true,width=\linewidth]{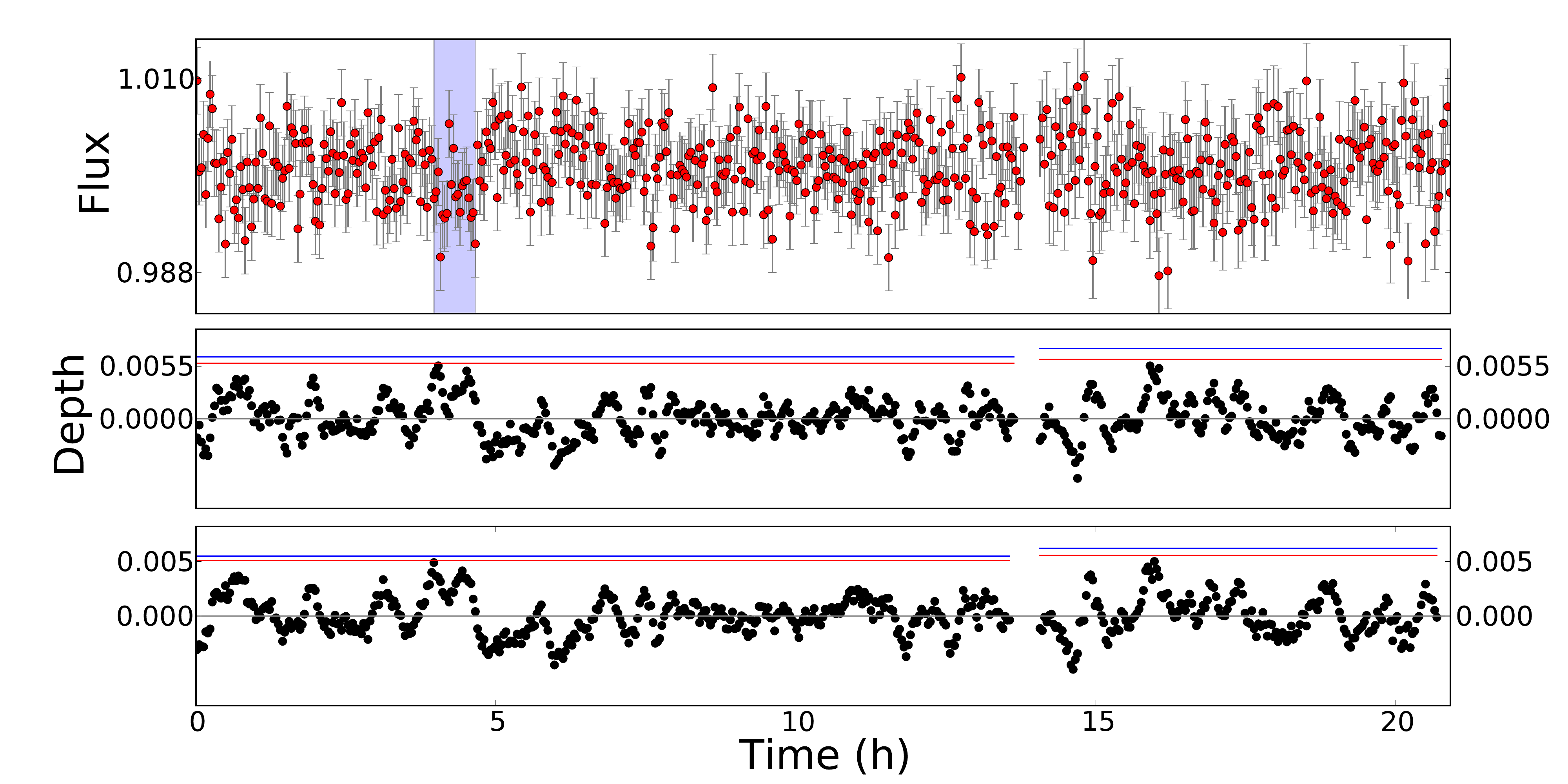}
\end{subfigure}
\begin{subfigure}{.44\textwidth}\caption{2M1516}
\centering
\includegraphics[trim={0.2cm 1.3cm 0 0.95cm},clip=true,width=\linewidth]{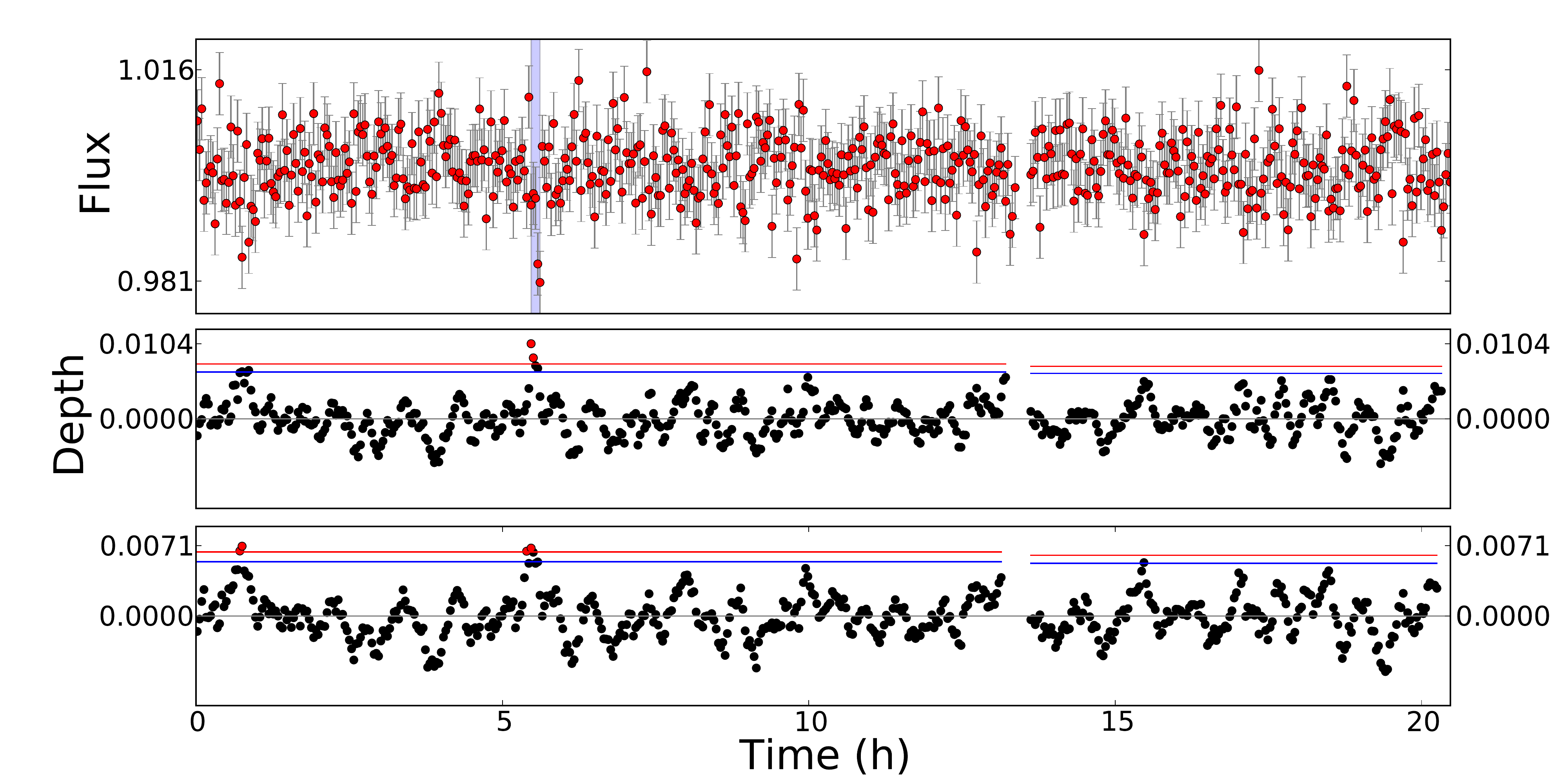}
\end{subfigure}
\begin{subfigure}{.44\textwidth}\caption{2M2208}
\centering
\includegraphics[trim={0.2cm 0.3cm 0 0.95cm},clip=true,width=\linewidth]{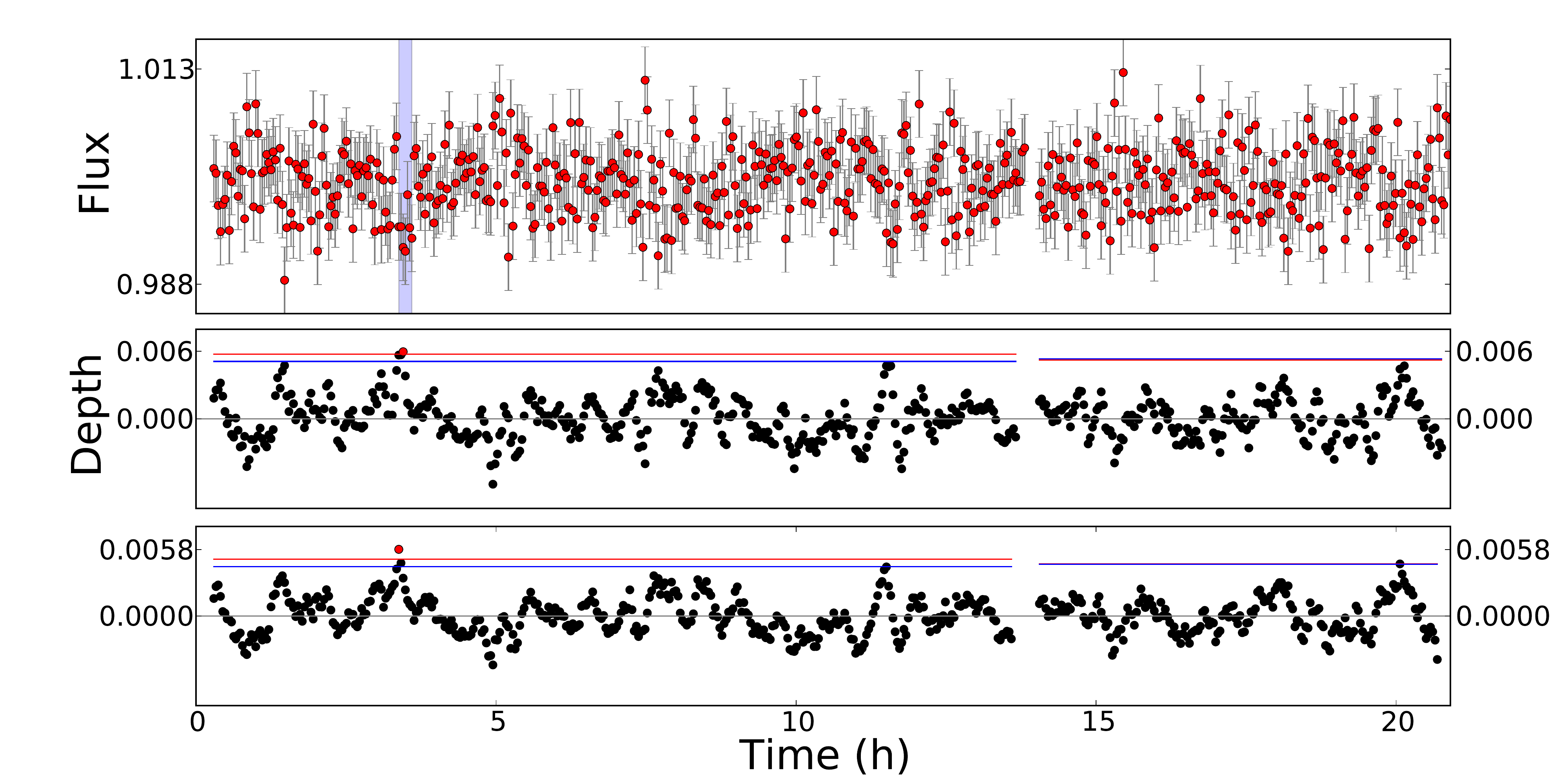}
\end{subfigure}
\begin{subfigure}{.44\textwidth}\caption{2M2228}
\centering
\includegraphics[trim={0.2cm 0.3cm 0 0.95cm},clip=true,width=\linewidth]{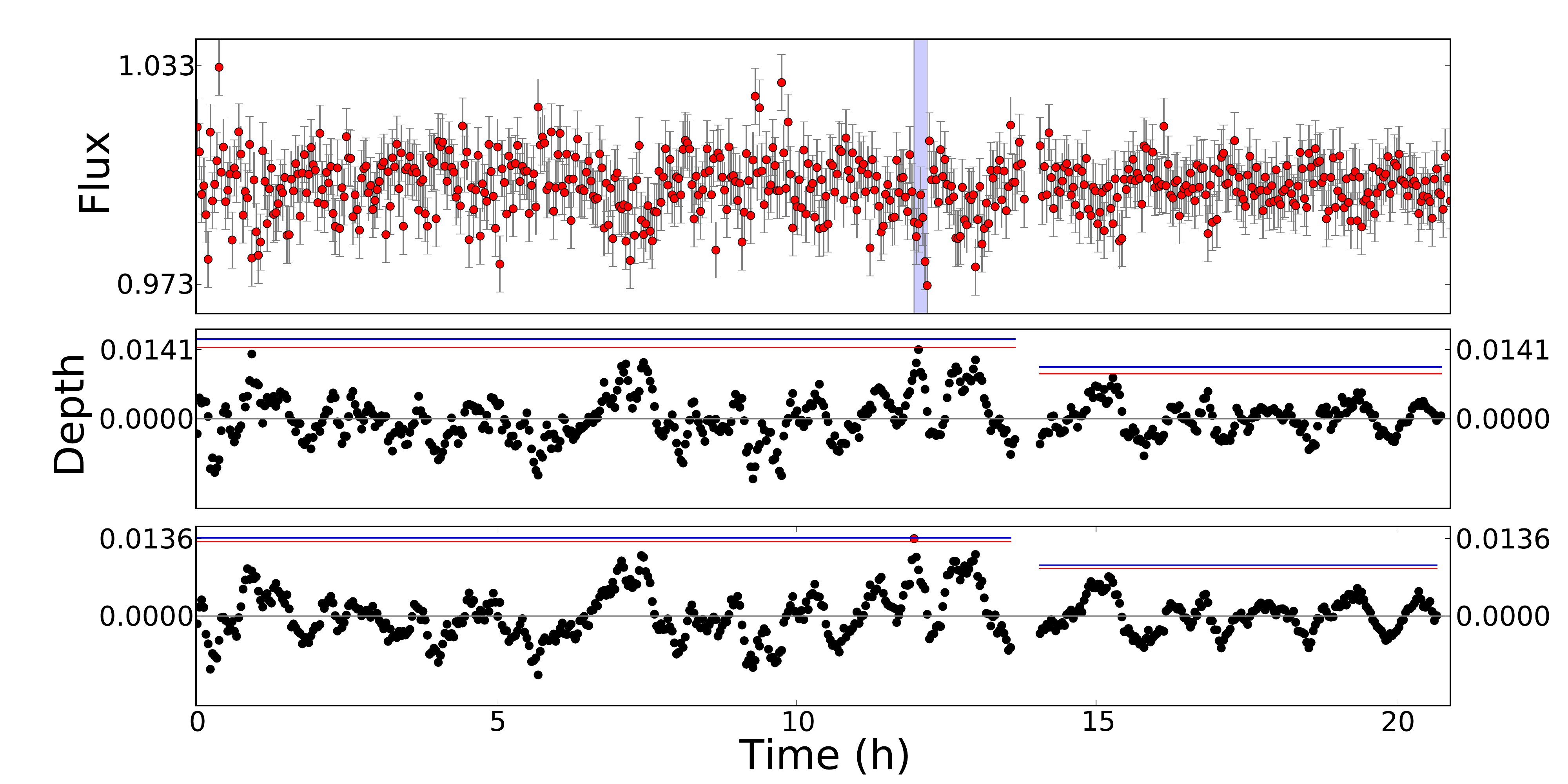}
\end{subfigure}
\caption{Six lightcurves showing candidate transit-like events.  Below each of the timeseries we plot the step by step depth found by the top-hat in black, for two two assumed width: $n=5$ (top) and $n=7$ (bottom). Other widths are not shown here. The red horizontal line indicate the three rms threshold, above which we count a detection. The blue horizontal line correspond to depths that have a depth producing an empirically determined SNR with a 50\% confidence for detection (i.e., derived from Figure \ref{fig:SNRhistograms}).}
\label{fig:metchev}
\end{figure*}

We insert and retrieve planetary transits into our light series by using a simplified model. A transiting planet produces a box-like event, which can be approximated with a {\it top-hat} model, essentially consisting of a vertical drop in flux, of a depth $D$ and for a duration (or width), $W$. These observable parameters can then be converted into physical quantities. For simplicity, we will neglect any limb darkening effect (which are irrelevant in the mid-infrared \citep{Claret:2011rm}, and unknown for brown dwarfs). We also neglect the typically brief ingress and egress durations, as is often done for detection algorithms \citep[][]{Kovacs:2002lr,Collier-Cameron:2007pb}.

We search for transits in the following way. We start the top-hat model such that the first point is at the beginning of the timeseries and assume a certain width. We compute the average flux within transit, and normalise the flux outside of the transit. The depth is the difference between both values. We then step the model by one point forward and repeat the procedure until we reach the end of the timeseries. This results in a temporal series of depth measurements. We calculate the mean and the rms of this series. Our criterion for a possible transit detection is when the transit depth reaches a value larger than three times the rms, above the mean depth value.
The entire procedure is repeated, assuming six different values for the width (5, 7, 10, 13, 16, and 20 points wide, corresponding to the range of actual transit durations in Table~\ref{tab:example}). 

Figure \ref{fig:lightcurvesmodels} shows examples of a few light curves that have inserted transits  and through which we ran the top-hat model, to identify interesting locations. We provide a range of scenarios: true detection, false positive, and false negative. We also display how the depth varies as a function of temporal location, for the width yielding the clearest signal. We show that we can identify individual events, which in some cases would have allowed us to identify both transit signals caused by a planet orbiting in less than 20.9 hours (or equivalently, that of two planets, each producing one event).

Although our procedure recovers $D$ and $W$ only, we insert transits using a range of physical parameters. This range will later be used to estimate our limits on the occurrence rate of planets orbiting brown dwarfs, described in Section~\ref{sec:occurence}. The orbital period $P$, its inclination from the plane of the sky $i_{\rm p}$, the transit phase $\phi_{\rm 0}$, and the planet radius $R_{\rm p}$ were randomly sampled. 

The orbital period was sampled linearly in $\log$ from about 0.23 to 3 days, where the lower limit was derived from assuming a minimum orbital separation of two times the Roche limit, $a_{\rm Roche} = 1.26 R_\star(\rho_\star/\rho_{\rm p})^{1/3}$ where we set the planet density $\rho_{\rm p}$ equal to the mean density of Earth by adopting $\rho_\oplus = 5.51$ g cm$^{-3}$. 
We drew the inclination angle sampling uniformly in $\cos{i_{\rm p}}$ with $i_{\rm p} = 0^{\circ}$ to $90^{\circ}$. $\phi_0$ was sampled linearly between 0 to 1. For our purposes of investigating Earth-sized planets, we also sampled planetary radii uniformly, between 0.25 and 3.25 R$_\oplus$. 

Every time a set of parameters is drawn we verify the planet's impact parameter and select only those that are in a transiting configuration. We stop when we reach 50\,000 transiting systems (which is on average 8.6\% of the total number of systems). We then randomly assign these systems to one of the 44 {\it Spitzer} lightcurves, and insert the transits where they fall by subtracting the depth $D$ for a number of points equivalent to the width $W$ (for the edge points of the transits, we scale the depth linearly by the fraction of the cadence that the in-transit duration covers). Finally we run our detection algorithm. A large fraction of the 50\,000 synthetic planets have a phase such that a transit happens during the {\it Spitzer} observations. Most of them are then identified by our search algorithm, with the detection rate falling quickly for systems with large orbital periods or very small planetary radii. We grouped the 50\,000 planets into three categories:
\begin{itemize}
\item the inserted transit falls outside the lightcurve;
\item the inserted transit is not recovered (but falls inside the lightcurve);
\item the inserted transit is properly recovered.
\end{itemize}
13.7\% fall in the first category, 11.5\% are not detected, and 74.8\% are properly detected. However, our algorithm gave a 37\% rate of falsely detected systems.  A fraction of this value arises from six transit-like candidates in the original 44 {\it Spitzer} lightcurves (see Sec.~\ref{sec:search}). 

We plot the number of detections and non-detections into Figure~\ref{fig:SNRhistograms} as a function of the signal to noise of the signal that had been inserted, as defined in eq.~\ref{eq:snr}.
For each {\it Spitzer} timeseries, we find the signal to noise ratios for which we have a 50\% and a 95\% probability to recover a correct transit. Those values are recorded in Table~\ref{tab:SNR50and95} where we also include what planet radii they correspond to (assuming an impact parameter of $b = 0.5$ and an orbital period equal to the average coverage of the timeseries, $P = 20.9$h, which translates to a width of about $W = 28$ min, or $n = 13$).
The 50\% recovery rate corresponds to an average signal to noise of 3.5, supplying support for the value we provided in Section~\ref{sec:signal} and Fig.~\ref{fig:rmshistograms}. We adopt this number as verification criterion for identifying transit-like signals from the data.


\section{Retrieval of Kepler-42b and TRAPPIST-1b}\label{sec:keplertrappist}

We take this paper as an occasion to compute an order of magnitude occurence rate for Earth-sized planets transiting objects at the bottom of the main sequence. We will later extrapolate that number to brown dwarfs and compute approximate yield estimates. We looked into the {\it Kepler Input Catalog}, and removed giant stars by applying colour-cuts following \citet{Dressing:2013xy} (J-K < 1.0 and Kepmag > 14). We then restricted ourselves to objects with Kepmag < 17.5, with $r'-K$ > 4.34, a colour compatible with ultra-cool dwarfs \citep[M/H = 0, age = 8 Gyr]{Baraffe:1998ly}. We obtain 107 stars, two of which have confirmed planets: Kepler-42 \citep{Muirhead:2012fk} and Kepler-445 \citep{Muirhead:2015lr}. These studies collected additional data that indicate that both systems orbit stars that are not ultra-cool dwarfs, but mid-M dwarfs instead. It is likely that the other members in the sample suffer similarly. Nevertheless, we keep this sample as the coldest within the {\it Kepler} data and therefore the most representative to the systems we seek.

Kepler-445's innermost planet would have less than 30\% chance of producing one transit within the duration of our {\it Spitzer} timeseries, so we conservatively discard it for our estimate. Kepler-42b has a $7\%$ geometric probability of transit. TRAPPIST-1b has a $5\%$ transit probability and the system was identified amongst 50 who have not all been fully monitored yet. The successful detection of these two transiting systems amongst a total of 157 surveyed combined with a mean chance of transit of $6\%$ suggests that the occurrence rate of Earth-like planets is likely on the order of $22\%$, within a two-day orbital period.

We performed a slightly more advanced analysis to confirm this number. We assumed a period distribution linear in $\log$ between the orbital periods of 0.55 and 6 days (so as to englobe most of the planets contained in the TRAPPIST-1, Kepler-42 and Kepler-445 systems, and ignoring the period box where we are the most sensitive). We modelled multiple sample of 157 stars, whose masses were randomly drawn from a uniform distribution in mass bounded by 80 and 150 M$_\odot$. We then computed which occurence rate was necessary to obtain the identification of two systems, using Poisson statistics. This yielded $\eta = 27\%$ (including Kepler-445, this number rises to $45\%$). We therefore conclude that the occurence is in the region of 20-30\%.

We extend our simulations to investigate whether we could detect Kepler-42 \citep{Muirhead:2012fk} and TRAPPIST-1 \citep{gillon:2016gh}. Both systems involve three Earth-sized planets at varying orbital periods:

\begin{itemize}
\item Kepler-42: radii of 0.73, 0.78, 0.57 R$_\oplus$ with periods of 0.453, 1.1214, 1.865 days respectively
\item TRAPPIST-1: radii of 1.11, 1.05, 1.16 R$_\oplus$ with periods of 1.51, 2.42, 18.202 days respectively
\end{itemize}

In order to determine the suitability of {\it Spitzer}'s photometry to detect these systems, we only considered the innermost planet of these systems (the planets with periods of 0.453 days and 1.51 days for Kepler-42b and TRAPPIST-1b respectively, i.e. 0.73 R$_\oplus$ and 1.11 R$_\oplus$), and simulate transit retrieval per each of the 44 lightcurves. For both planets, we sampled all other physical parameters (inclination $i_p$ and phase $\phi_0$) as described earlier. 
We still assume a $M_\star = 60M_{\rm Jup}$ and $R_\star = 0.9R_{\rm Jup}$ brown dwarf. 
We then ran the simulated data through our retrieval algorithm.

Our results indicate that for Kepler-42b, 10/44 lightcurves yield $>95\%$ detection rate with none of them giving $>99\%$ detection. For TRAPPIST-1b, 33/44 lightcurves give $>95\%$ detection rate with 15/44 yielding $>99\%$ retrieval, results which are markedly better due to the planet being larger. To summarise, $75\%$ of our 44 Spitzer series provide a $95\%$ confidence for detecting a 1.11 R$_\oplus$ planet with a period of 1.51 days, indicating a remarkable sensitivity to transiting Earth-sized planets.

\begin{figure}
\centering
\includegraphics[width=0.45\textwidth]{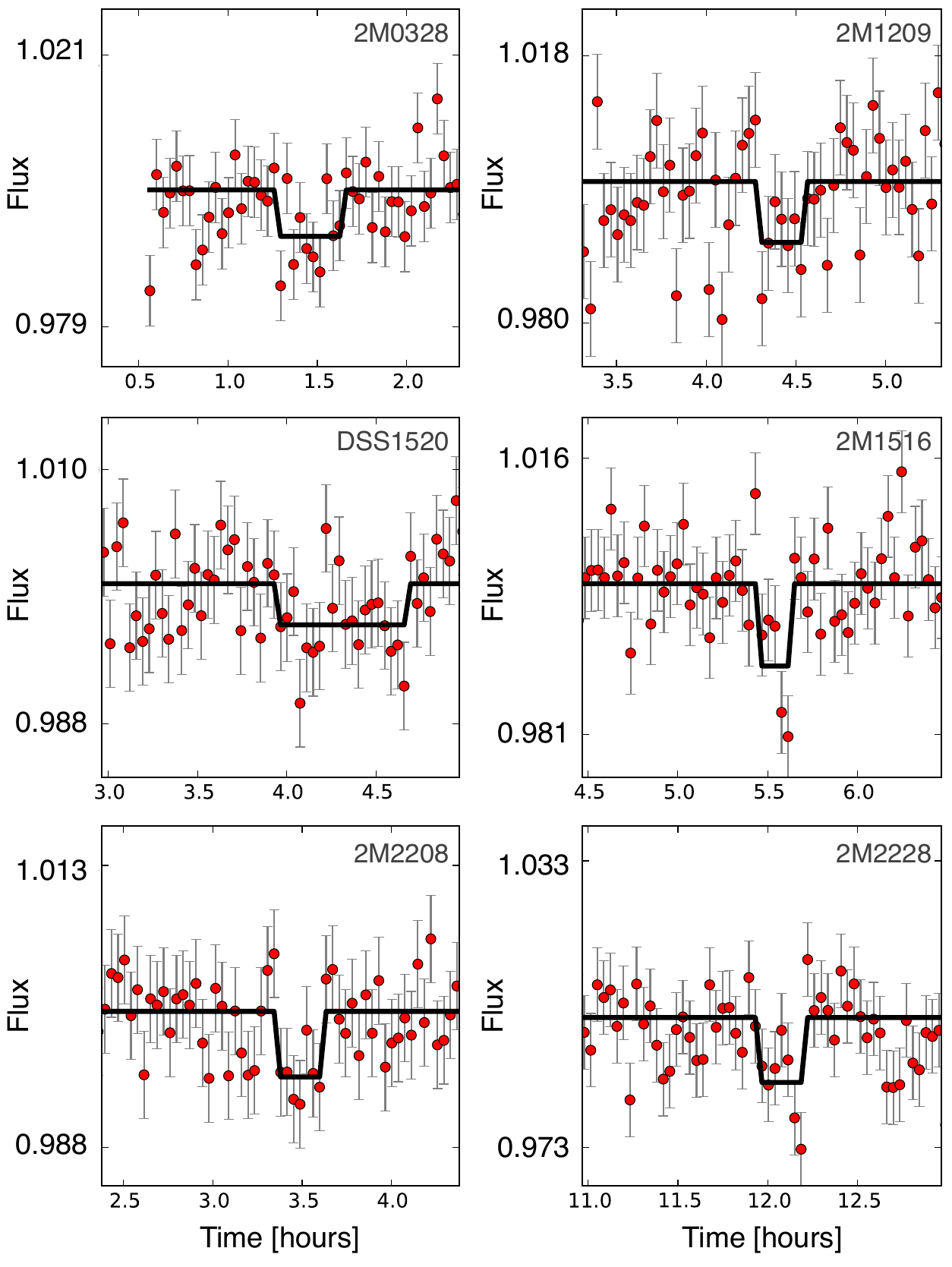}
\caption{Close up view on the transit-shaped shaped signal for the six candidates that passed our selection criteria.}
\label{fig:metchevzoomed}
\end{figure}

\begin{table*}
\centering
\caption{Transit candidates retrieved using our 3 rms depth criterion that have SNR greater than the 50\% threshold (i.e., corresponding to the objects in Figure \ref{fig:metchev}). The transit epoch corresponds to the front-end position of our top-hat model that gives detection. For each object, the width $W$ corresponding to the maximum SNR (as shown in the third column) out of the six widths we tried is presented, along with the respective depth $D$ for that width. The radii R$_p$ of the potential planets are calculated from the depths assuming R$_\star$ = 0.9 R$_{\rm Jup}$. We also show the magnitudes of the targets in 2MASS J and K bands, and in the {\it Wise} 1 and 2 channels, which are similar to {\it Spitzer's}.}
\label{tab:candidates}
\begin{tabular}{lccccccccc}
\hline
\hline
Object & Transit & Max. & $W$ & $D$ & R$_p$ & $J$ & $K$ &W1 & W2\\
 & epoch {[h]} & SNR & [min] & {[\%]} & {[R$_\oplus$]} &{[mag]} & {[mag]}& {[mag]}& {[mag]}\\
\hline
2MASS 0328 & 1.6 & 4.0 & 21.8 & 0.77 & 0.96 &16.69 &14.87		&14.13 &13.60\\
2MASS 1209 & 4.3 & 3.2 & 15.3 & 0.87 & 1.02 &15.91 & 15.06		&14.66 & 13.48\\
2MASS 1516 & 5.5 & 4.0 & 10.9 & 1.04 & 1.12 & 16.85 & 15.08	&14.14 &13.40\\
SDSS 1520 & 4.0 & 4.1 & 43.6 & 0.36 & 0.66 &15.54 & 14.00		&13.44 &12.92\\
2MASS 2208 & 3.4 & 3.7 & 15.3 & 0.58 & 0.84 &15.80 & 14.15 	&13.38 &12.91\\
2MASS 2228 & 12.0 & 4.3 & 15.3 & 1.36 & 1.28 & 15.66 & 15.30	&15.23 &13.33\\
\hline
\end{tabular}
\end{table*}

\section{Search for transit-like signals}\label{sec:search}

We search each of the {\it Spitzer} lightcurves for transit-like signals, using the retrieval procedure described above. Despite {\it Spitzer}'s precise photometry, we detect no obvious candidate for transiting exoplanets, only recovering a number of doubtful events. In 20 cases we obtained identified locations where the top-hat indicates a lower flux for at least one of our six model widths. In 14 systems, the lowest depths recovered have a SNR high enough for a 50\% likelihood that the event is real (as produced in Table~\ref{tab:SNR50and95}), although only a number of them are detected according to our criterion. The overlap includes 6 candidates which have a signal that is both over the 3 rms depth threshold and over the 50\% SNR threshold. We present these candidates in Figure~\ref{fig:metchev}. None of these events repeat, implying periods larger than 21 hours. None of these reaches a 95\% confidence detection and are therefore likely false positives. 
In Table~\ref{tab:candidates} we present the details of these detected best candidates. The transit epoch is displayed as the time of the beginning of the transit model which gives the detection. Since we ran six different top-hat model widths through each light curve and thus in some of these cases the same transit-like signal is detected by different widths, in order to choose the one model (width $W$ and corresponding depth $D$) we opted for the one yielding the highest SNR.

\begin{figure*}
\centering
\includegraphics[width=0.85\textwidth]{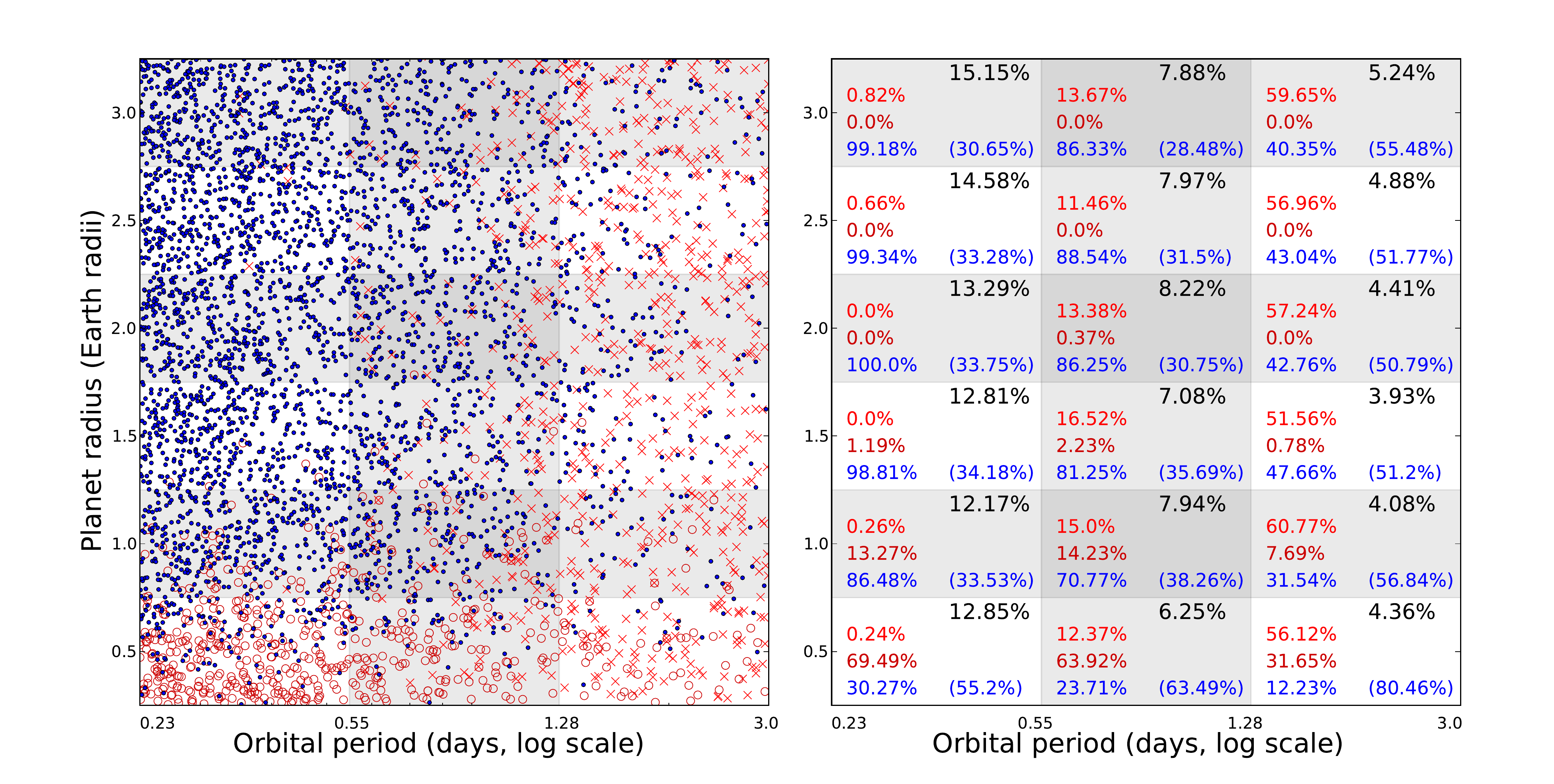}
\caption{{\bf Left:} Map of transiting planets versus planet radius and orbital period. Blue filled in dots represent transiting planets that are detected, hollow red circles represent transiting planets that are captured on the light curve but are not detected, and red x's represent transiting planets that are missed by the time series. {\bf Right:} The corresponding transit fractions and detection statistics per box: the top right corner indicates the probability of transit, while the three percentages on the left side of each box represent, from top to bottom, the fraction of transits missed by the observation (red x's), the fraction of transits undetected using our methods (hollow red circles), and the fraction of detected transits (blue dots). The percentage in parenthesis in the bottom right corner of each box represent the rate of false positives.}
\label{fig:radiusperiodmap}
\end{figure*}

\section{Estimates of maximum occurrence rate}\label{sec:occurence}

We use the population of 50\,000 synthetic transiting planets that we described in Sec.~\ref{sec:retrieval}, to measure upper limits on the occurence rate of planets orbiting brown dwarfs. We divide the planet radius and orbital period domain into a grid of 6 rows (equally spaced in R$_{\rm p}$ from 0.25 to 3.25 R$_\oplus$) and 3 columns (equally spaced in $\log P$, from $P = 0.23$ to $3$ days), consisting of 18 boxes. Inside each box we compute:
\begin{itemize} 
\item the mean transit probability,
\item the fraction of transits missed (not in the light curve),
\item the fraction of properly recovered transits.
\item the fraction of false positives.
\end {itemize}
The results are plotted in Figure \ref{fig:radiusperiodmap}, for a subset of the total population to avoid over-crowding (10\% of the total).

\begin{figure}
\centering
\includegraphics[width=0.45\textwidth]{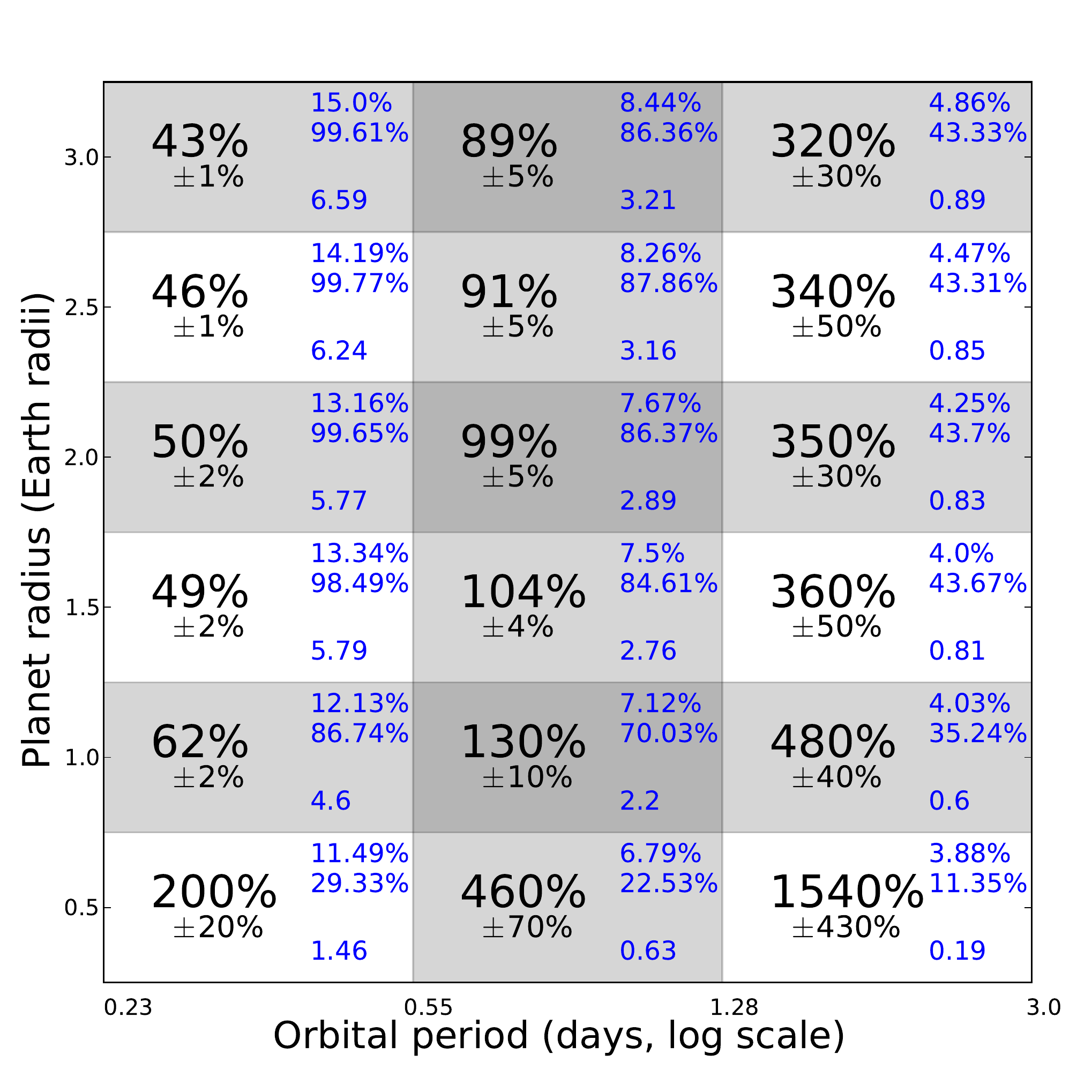}
\caption{Info-graphic showing the fractions of brown dwarfs that would have a planet (in black, center of each box), in order to be consistent with having no detections for 44 observations, 95\% of the time. The top right corner of each box gives the probability of transit (top-most) and detection chance of those transits as derived from sampling. The bottom right value gives the number of detections we would expect for $\eta = 1$, of our 44 brown dwarfs.}
\label{fig:predictedfractionsmap}
\end{figure}

We then proceed to estimate the occurrence rate, which we denote as $\eta$. By setting $\eta = 1$ and multiplying it by the transit probability $p$, and the fraction of properly recovered transits $f$, we can get a rapid idea of how constraining the observations are. We place those numbers in Fig.~\ref{fig:predictedfractionsmap} (bottom right of each box). For instance, in the top-left box, 6.6 planets would be expected. Since we have no detection, we can only estimate upper limits in each box, and for a combination of boxes. Each of the values that we provide are 95\% confident. $\eta$ is calculated by considering the probability of detecting a transit given a random brown dwarf and demanding that the chance of having 44 observations with no detections in the box is 5\%. Eq.~\ref{eq:etaestimate} encapsulates this calculation by rearranging to solve for $\eta$:
\begin{equation}\label{eq:etaestimate}
0.05 = (1-\eta{f}{p})^{n}
\end{equation}
where $f$ is the fraction of properly recovered transits, $p$ is the probability of transits, and $n$ is the number of targets in our sample (44). We gather the results in Fig.~\ref{fig:predictedfractionsmap}. In our example of the top-left box, we find $\eta < 43\pm1\%$, which is the tightest limit using this portion of parameter space. The uncertainty on $\eta$ is derived by dividing the 50\,000 transits into ten equal sets, repeating the calculations, and computing the rms of the ten trials. Note that  $\eta > 100\%$ means that the brown dwarfs would have more than one planet each on average in a particular box.

Several trends are evident from the values presented in Figure \ref{fig:predictedfractionsmap}. The sharp increase in $\eta$ for the bottom row (planets from 0.25 to 0.75 R$_\oplus$) is mostly caused by the fact that such small planets are extremely hard to detect, even for the relatively precise photometry obtained of [3.6] and [4.5]. The probability of transit also slightly reduces with planet radius; smaller planets are slightly less likely to transit for a given orbital inclination. This creates a drop in $p$ by a few percentage points where otherwise detection fractions are nearly 100\%. The very large values of $\eta$ for the longer orbital periods reflects two important effects: the orbital period exceeds the timeseries' coverage (thus there is an increased chance that the transit is not captured), and the geometric probability of transit plummets with orbital period.

We will now combine different regions of parameter space and produce a number of scenarios. We also compute how many brown dwarfs need to be surveyed in order to have a 95\% chance of obtaining a detection. In this exercise we assume that new targets will follow the distributions described in Fig.~\ref{fig:rmshistograms}, when in fact one could bias the sample to earlier spectral types and brighter targets.

\subsection{Between 0.75 and 3.25 R$_\oplus$}

We estimate the overall $\eta$ for planets with sizes between 0.75 and 3.25 R$_\oplus$ over all three period boxes. Our 95\% confident upper limit is $\eta < 92 \pm 2 \%$. Restricting ourselves to the two columns at shorter periods, we obtain $\eta < 67 \pm 1 \%$. 
If in reverse, there are no planets in the inner period column, we get $\eta < 159 \pm 4 \%$ for just the two outer columns, and $\eta < 102 \pm 3 \%$ simply for the central column.

Now, we assume there is a flat $\eta = 10\%$. In order to make one detection (with 95\% confidence), we need to observe 413 targets with brightness and variability similar to our sample of 44, for a 20.9 hour monitoring, and 365 targets if we instead observe for 42 hours. If we assume no planets for the inner period box and a 10\% rate for the two outer boxes, those numbers rise to 717 and 542 respectively.

\subsection{Earth-sized planets}

We estimate $\eta$ for planets with sizes between 0.75 and 1.25 R$_\oplus$, likely to be rocky \citep[][]{Rogers:2015qy}, for which we also retain a good rate of recoverability. On the ensemble of three period boxes, we place a 95\% confident upper limit of $\eta < 120 \pm 4 \%$. Restricting ourself to the inner two boxes, we find $\eta < 87 \pm 3\%$.

Assuming $\eta = 10\%$ spread over the three boxes, we expect detectable Earth-like transiting planets for 0.63\% of brown dwarfs, over a 20.9h timeseries. This value increases to 0.72\%, assuming we extend the photometric coverage to 42 hours. For a 20.9h and a 42h monitoring per brown dwarf, we would need to survey 476 and 417 systems respectively, in order to have a 95\% confidence for a successful detection.

If we adopt the results of the Kepler-42b and TRAPPIST-1b systems and set $\eta = 27\%$ for the three Earth-sized boxes as derived in a previous section, the number of systems we need to survey for a 95\% chance of detecting one  planet drops to just 175 (of which 44 have already been monitored), given a 20.9h monitoring. Doubling the observation time to 42h monitoring reduces this number to 153 brown dwarfs, with brightness, and variability similar to the WoW sample.

\section{Discussion}\label{sec:discuss}

We have searched for transiting planets signals within photometric timeseries obtained on 44 brown dwarfs by the Weather on other Worlds programme \citep{Metchev:2015rf}, that used the {\it Spitzer} space telescope. After studying  our ability to recover transits with synthetic signals, we identify six transit event candidates. They have depths at the threshold of our detection criteria and are therefore likely to be false positives. We assumed the lightcurves contained no transits and generated a size/period map of upper limits on the occurence rates of planets orbiting brown dwarfs.

\begin{figure}
\centering
\includegraphics[width=0.45\textwidth]{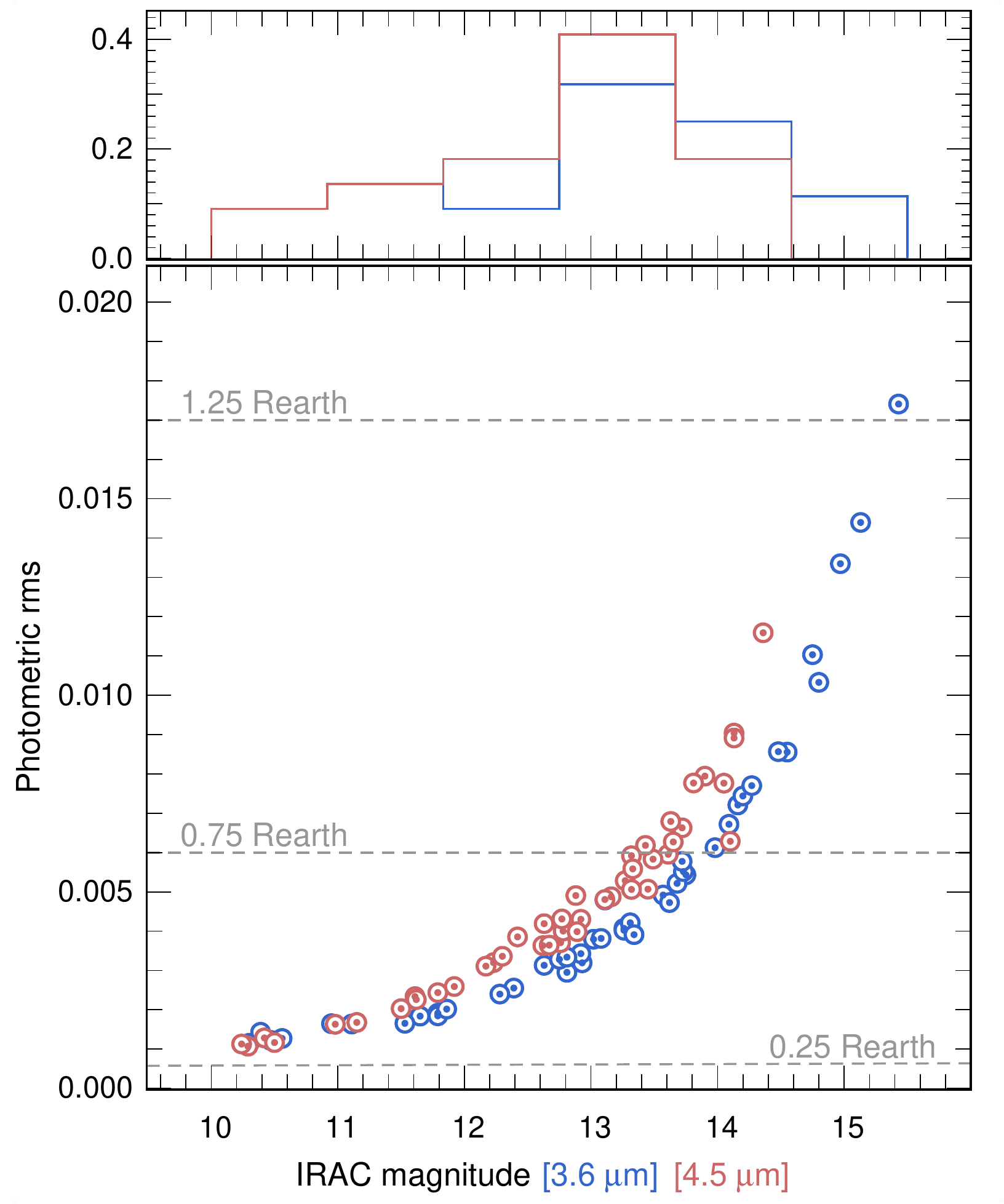}
\caption{{\it Spitzer}'s photometric precision as a function of the magnitude of the source in IRAC's channel 1 and 2 (3.6 and 3.5 $\mu$m). We show a histogram of the magnitudes in the top panel, and indicate to which planet size, photometric precision correspond to (for a SNR of 3.5).
}
\label{fig:irac}
\end{figure}

Due to the limited number of targets, and short observation time, our results are not very constraining, but can inform us about how much more of an observational effort can be made to discover planets orbiting brown dwarfs, using the transit method. The discovery of an Earth-sized planet transiting a nearby and relatively bright brown dwarf would provide an exquisite target for the detailed characterisation of a terrestrial extrasolar world, one that will be similar to Earth in some aspects (size, temperature...), but different in others (irradiation, tidal locking...). With the current sample and assuming a 27\% occurence rate (as implied by the discoveries of TRAPPIST-1 \citep{gillon:2016gh} and Kepler-42 \citep{Muirhead:2012fk}), we estimate that 175 brown dwarfs need to be monitored for 21 hours, of which 44 have already been done. This sample of brown dwarfs, that composed the WoW survey, is centred on the L-T transition, the focus of most of brown dwarf variability. In addition many of these targets were not selected for their brightness (see Fig.~\ref{fig:irac}). A careful sample selection concentrating on brown dwarfs with magnitudes brighter than 12 (in both {\it Spitzer} channels) would likely lower this required number down. A doubling of the observing time only marginally helps transit recovery (also because we used a `single transit' detection approach).

Finding one Earth-sized planet transiting a nearby brown dwarf, would provide an exquisite target for the detailed characterisation of a terrestrial extrasolar world using transmission \citep[like for TRAPPIST-1; ][]{gillon:2016gh,de-Wit:2016fk}, and emission spectroscopy. In addition, the discovery of multiple planet orbiting brown dwarfs may help answer fundamental questions in Astrobiology, yielding precious information on the physical requirement and timescale for the emergence of life. 

Stars evolve, which changes the irradiation received by a planet, and therefore its ability to retain liquid water at the surface. This implies that many planets remain within the habitable zone only for a while. Some enter it at a late stage in their history. This argument has been used by \citet{Ramirez:2016qy}. They argue that as solar-like stars leave the main sequence and evolve to the stage of giants, the habitable zone sweeps worlds that were previously too cold. We would counter-argue that as a star becomes a giant, it becomes harder to distinguish the planet from the star due to a worsening flux and size ratio. However the main idea, that of evolution remains valid, and very useful.

Brown dwarfs evolve all the time. They shrink and cool down \citep[eg.][]{Burrows:1997fj,Baraffe:2003gf} constantly. Planets that were previously not within the habitable zone will eventually enter it, sometimes Gyrs after their formation. Identifying such planets, and finding traces of life inside their atmospheres would inform us that Archean conditions are not necessary for the emergence of life. In addition, the time a given planet will remain habitable will vary as a function of the mass of the brown dwarf host. Once a statistically significant sample of habitable worlds has been gathered orbiting a variety of brown dwarfs, it will become possible to measure a timescale for the emergence of biology from the amount of time spent being habitable and the relative number of planets.


\section*{Acknowledgements}
\addcontentsline{toc}{section}{Acknowledgements}

We are grateful, and thank our anonymous referee for taking the time to read and assess our paper.
MYH would acknowledge and thank Yanqin Wu for supporting this research financially. AHMJT was at the time a fellow of the Centre for Planetary Sciences, hosted at the University of Toronto. MG is a Research Associate at the Belgian Scientific Research Foundation (F.R.S-FNRS). MYH and AHMJT acknowledge the many discussions and friendly collaborations that existed and allowed this project to fruition with Yanqin Wu, Daniel Tamayo, Hanno Rein, Diana Valencia and Kristen Menou. In addition, our thoughts about planets transiting brown dwarfs were influenced via interactions with Frank Selsis, Sean Raymond, \'Emeline Bolmont, J\'er\'emy Leconte, Brice-Olivier Demory, Josh Winn, Stanimir Metchev and Adam Burgasser. We are grateful to Stanimir Metchev and to Aren Heinze for sending us the lightcurves resulting from their analysis of the WoW programme.

This publication makes use of data products from the Wide-field
    Infrared Survey Explorer \citep{Cutri:2013fj}, which is a joint project of the University of
    California, Los Angeles, and the Jet Propulsion Laboratory/California
    Institute of Technology, funded by the National Aeronautics and Space
    Administration.
    
We also used data products from the Two Micron All Sky Survey \citep{Skrutskie:2006kx}, which is a joint project of the University of Massachusetts and the Infrared Processing and Analysis Center/California Institute of Technology, funded by the National Aeronautics and Space Administration and the National Science Foundation.





\bibliographystyle{mn2e}
\bibliography{../1Mybib.bib}





\bsp	
\label{lastpage}
\end{document}